\documentclass{article}

\usepackage[margin=1.25in]{geometry}
\usepackage{microtype}
\usepackage{graphicx}
\usepackage{subfigure}
\usepackage{booktabs} 

\usepackage{amsmath}
\usepackage{amssymb}
\usepackage{amsthm}
\usepackage{cases}

\newtheorem{theorem}{Theorem}

\newtheorem{proposition}{Proposition}
\usepackage[utf8]{inputenc} 
\usepackage[T1]{fontenc}    
\usepackage{url}            
\usepackage{amsfonts}       
\usepackage{nicefrac}       
\usepackage{sidecap}
\usepackage{tikz}
\usepackage{subfiles}
\usepackage{times}
\usepackage{helvet}
\usepackage{courier}
\usepackage{url}
\usepackage{enumerate}
\usepackage{natbib}
\usepackage{algorithm}
\usepackage{algorithmic}
\usepackage{authblk}
\usepackage{natbib}

\usetikzlibrary{decorations.pathreplacing}

\usepackage{hyperref}

\renewcommand{\cite}[1]{\citep{#1}}

\title{Submodular Cost Submodular Cover with an Approximate Oracle}
\author[1,*]{Victoria G. Crawford}
\author[2]{Alan Kuhnle}
\author[1]{My T. Thai}
\affil[1]{University of Florida}
\affil[2]{Florida State University}
\affil[*]{vcrawford01@ufl.edu}
\date{}                     
\begin{document}

\maketitle

\begin{abstract}
In this work, we study the Submodular Cost Submodular Cover problem,
which is to minimize the submodular cost required to ensure that the submodular
benefit function exceeds a given threshold.
Existing approximation ratios for the greedy algorithm
assume a value oracle to the benefit function.
However, access to a value oracle is not a realistic assumption for many applications of this problem,
where the benefit function is difficult to compute.
We present two incomparable approximation ratios for this problem
with an approximate value oracle and demonstrate that the ratios take on
empirically relevant values through a case study with the Influence Threshold problem in online
social networks.
\end{abstract}

\section{Introduction}
Monotone\footnote{For all $A\subseteq B\subseteq S$, $f(A)\leq f(B).$} submodular
set functions
are found in many applications in machine learning and data mining \cite{kempe2003,lin2011,wei2013,singla2016}.
A function $f:2^S \to \mathbb R$ defined on subsets of a ground set $S$ is submodular if for all
$A\subseteq B \subseteq S$ and $x\notin B$,
\begin{align*}
  f(A\cup\{x\})-f(A) \geq f(B\cup\{x\})-f(B).
\end{align*}
The ubiquity of submodular functions
ensures that the optimization of submodular functions has received much attention
\cite{Nemhauser1978,wolsey1982}.
In this work, we study
the Submodular Cost Submodular Cover (SCSC)
optimization problem, originally introduced
by \citeauthor{wan2010} (\citeyear{wan2010}) as a generalization of \citeauthor{wolsey1982} (\citeyear{wolsey1982}).
SCSC is defined as follows.

\paragraph{Submodular Cost Submodular Cover (SCSC)}
Let $f,c:2^S\to \mathbb{R}_{\geq 0}$\footnote{We also assume that $f(\emptyset)=0$,
and $c(X)=0$ if and only if $X=\emptyset$.} be monotone submodular functions
defined on subsets of a ground set $S$ of size $n$.
Given threshold $\tau \leq f(S)$, find $\text{argmin}\{c(X)| X\subseteq S, f(X)\geq\tau\}$.

Applications of SCSC include influence in social networks \cite{goyal2013,kuhnle2017},
data summarization \cite{mirzasoleiman2015,mirzasoleiman2016},
active set selection \cite{norouzi2016}, recommendation systems \cite{guillory2011},
and monitor placement \cite{soma2015,zhang2016detecting}.

Existing approximations \cite{wolsey1982,wan2010,soma2015} to the NP-hard
SCSC problem assume value oracle access to $f$, meaning that $f$ can be queried
at any subset $X\subseteq S$.
Unfortunately, for many emerging applications of submodular functions, the function $f$
is difficult to compute. Instead of having access directly to $f$, we
may query only a surrogate function $F$ that is $\epsilon$-approximate
to $f$, meaning that for all $X\subseteq S$
\begin{align*}
  |f(X)-F(X)| \leq \epsilon.
\end{align*}
For example, $f$ may be approximated by a sketch \cite{badanidiyuru2012,cohen2014},
evaluated under noise \cite{chen2015,singla2016},
estimated via simulation \cite{kempe2003},
or approximated by a learned function \cite{balcan2012}.

If the surrogate function $F$ is monotone submodular, then we may use existing
approximation results for SCSC (see Appendix \ref{section:submodularF}).
However, it is not always the case
that the surrogate $F$ maintains these properties. For example, the approximate influence oracle
of \citeauthor{cohen2014} (\citeyear{cohen2014}) described in detail in Section \ref{section:experiments}
is non-submodular.
To the best of our knowledge, no approximation results currently exist
for the SCSC problem under a general approximate oracle.

\subsection{Our Contributions}
  We provide an approximation ratio for SCSC if the greedy algorithm
  (Algorithm \ref{algorithm:greedy})
  has a value oracle to an $\epsilon$-approximate function to $f$, provided that the
  smallest marginal gain $\mu$\footnote{See notation in Section \ref{section:definitions} for a definition.}
  of any element that was added to the greedy solution is sufficiently large
  relative to $\epsilon$ (Theorem \ref{ratio}).
  Our proof of Theorem \ref{ratio} is a novel adaptation of the charging argument developed by
  \citeauthor{wan2010} (\citeyear{wan2010}) for SCSC with integral-valued $f$,
  and has potential to be used for other versions of SCSC
  where $f$ cannot be evaluated for all $X\subseteq S$.
  If the oracle error $\epsilon = 0$, our ratio nearly reduces to existing ratios for
  SCSC \cite{wolsey1982,wan2010,soma2015}.

  We provide a second, incomparable approximation ratio for SCSC if the greedy algorithm
  has a value oracle to an $\epsilon$-approximate function to $f$, under the same conditions on
  the marginal gain as Theorem \ref{ratio} (Theorem \ref{ratio2}). In practical scenarios, the ratio of
  Theorem \ref{ratio} is sometimes difficult to compute or bound because it requires evaluation of $f$.
  In contrast, an upper bound on the ratio of Theorem \ref{ratio2} is easy to compute only having access to
  the surrogate $F$.
  If the oracle error $\epsilon = 0$, our ratio is a new approximation ratio
  for SCSC that is incomparable to existing ratios for SCSC \cite{wolsey1982,wan2010,soma2015}.

  We demonstrate that the ratios of Theorem \ref{ratio} and Theorem \ref{ratio2} take on
  empirically relevant values through a case study with the Influence Threshold (IT) problem under the
  independent cascade model on real
  social network datasets. This problem is a natural example of SCSC in which the
  function $f$ is the expected activation of a seed set and is
  $\# P$-hard to compute, so optimization must proceed with a suitable surrogate function $F$.
  We use the average reachability sketch proposed by \citeauthor{cohen2014} (\citeyear{cohen2014})
  for $F$, which is a non-submodular $\epsilon$-approximation of $f$.

\paragraph{Organization}
In Section \ref{section:relatedwork}, we present an overview of related work on
SCSC and submodular optimization with approximate oracles. Definitions used throughout the paper
are presented in Section \ref{section:definitions}.
The main approximation results (Theorems \ref{ratio} and \ref{ratio2}) are
presented in Section \ref{section:mainresults}.
We consider the special case where $F$ is monotone submodular in Appendix \ref{section:submodularF}.
Finally, in Section \ref{section:experiments}, we compute the values that the approximation ratios of Theorems
\ref{ratio} and \ref{ratio2} take on for Influence Threshold.


\subsection{Related Work}
\label{section:relatedwork}

\paragraph{Submodular Cost Submodular Cover (SCSC)}
Approximation guarantees of the greedy algorithm with value oracle access to
$f$ for SCSC have previously been analyzed \cite{wolsey1982,wan2010,soma2015}.

If $c$ is modular\footnote{$c$ is modular if for all $X\subseteq S$, $c(X)=\sum_{x\in X}c(\{x\})$.}
and $f$ is integral valued, then \citeauthor{wolsey1982} (\citeyear{wolsey1982})
proved that the approximation ratio of the greedy algorithm is
$\ln(\alpha)$, where $\alpha$ is the largest singleton value of $f$\footnote{$\alpha=max_{x\in S}f(\{x\})$}.
This is the best we can expect: set cover, which is a special case of SCSC,
cannot be approximated within $(1-\epsilon)\ln(n)$ unless
NP has $n^{\mathcal{O}(\log(\log(n)))}$-time deterministic algorithms \cite{feige1998}.

If $f$ is real-valued, \citeauthor{wolsey1982}
proved that the greedy algorithm has an approximation ratio of
$1+\ln(\alpha/\beta)$, where $\beta$ is the smallest non-zero marginal gain of adding any element to
the greedy solution at any iteration\footnote{See notation in Section \ref{section:definitions} for a definition.}.
In comparison, when the cost is modular and the value oracle exact ($\epsilon=0$) then the ratio that we provide in
Theorem \ref{ratio} reduces to $2+\ln(\alpha/\beta)$.

If $f$ is integral,
\citeauthor{wan2010} (\citeyear{wan2010}) proved that
the greedy algorithm has an approximation ratio of $\rho \ln(\alpha)$,
where $\rho$ is the curvature of $c$. \citeauthor{wan2010} developed a charging
argument in order to deal with the general monotone submodular cost function.
In the argument of \citeauthor{wan2010}, the cost of the greedy solution, $c(A)$,
is split up into charges over the elements of the
optimal solution $A^*$. This method of charging will not work for SCSC with an $\epsilon$-approximate oracle.
This is because the elements chosen by the surrogate $F$ do not necessarily exhibit
diminishing cost-effectiveness. But, our argument is inspired by that of \citeauthor{wan2010}.
Portions of our argument that share significant overlap with that of \citeauthor{wan2010} are made
clear and restricted to the appendix.
When $f$ is integral and the value oracle exact ($\epsilon=0$), the ratio that we provide in
Theorem \ref{ratio} reduces to $\rho(2+\ln(\alpha))$.

\citeauthor{soma2015} (\citeyear{soma2015}) generalized SCSC to functions on the integer lattice,
an extension of set functions.
\citeauthor{soma2015} proved that a decreasing threshold algorithm
has a bicriteria approximation ratio of
$(1+3\delta)\rho(1+\ln(\alpha/\beta))$ to SCSC on the integer lattice,
where $\delta < 1$ is an input.
When the value oracle is exact ($\epsilon=0$), the ratio that we provide in
Theorem \ref{ratio} reduces to $\rho(2+\ln(\alpha/\beta))$.

The special case of SCSC where $c$ is cardinality is the Submodular Cover (SC) problem.
Distributed algorithms \cite{mirzasoleiman2015,mirzasoleiman2016} as well as
streaming algorithms \cite{norouzi2016} for SC have been developed and their approximation
guarantees analyzed.

To the best of our knowledge, we are the first to study
SCSC with an approximate oracle to $f$.

\paragraph{Optimization with Approximate Oracles}
A related problem to SCSC is Submodular Maximization (SM) with a cardinality constraint\footnote{
Given a budget $\kappa$ and a
monotone submodular function $f$ defined on subsets of
a ground set $S$ of size $n$, find argmax$\{f(X):|A|\leq\kappa\}$.}.
SM with a cardinality constraint and an approximate oracle
has previously been analyzed \cite{horel2016,qian2017}.

\citeauthor{horel2016} (\citeyear{horel2016}) considered SM with a cardinality constraint
where we seek to maximize $F$ which is a
relative\footnote{See discussion in Section \ref{section:definitions} about $\epsilon$-approximation.}
$\epsilon$-approximation
to a monotone submodular function $f$.
Under certain conditions on the oracle error,
\citeauthor{horel2016} found that
the greedy algorithm yields a tight approximation ratio of
$1 - 1/e - \mathcal{O}(\delta)$.
On the other hand, \citeauthor{horel2016} proved for any fixed $\beta > 0$,
no algorithm given access to a $1/n^{1/2-\beta}$-approximate $F$ can have a
constant approximation ratio using polynomially many queries to $F$.
The results of \citeauthor{horel2016} can easily be translated into maximization of $f$ with an approximate oracle.
Another recent work for SM with a cardinality constraint and an approximate oracle
is \citeauthor{qian2017} (\citeyear{qian2017}), where a Pareto algorithm is analyzed.

A dual problem to SCSC is Submodular Cost Submodular Knapsack (SCSK)\footnote{
Given a budget $\kappa$ and
monotone submodular functions $f,c$ defined on subsets of
a ground set $S$ of size $n$, find argmax$\{f(X):c(A)\leq\kappa\}$.}.
In the absence of oracle error,
the approximation guarantees of SCSC and SCSK are connected \cite{iyer2013}.
In particular, \citeauthor{iyer2013} proved that
an $(\alpha,\delta)$-bicriteria\footnote{See Section \ref{section:definitions} for a discussion of bicriteria
approximation guarantees.} approximation
algorithm for SCSK can be used to get a $((1+\epsilon)\delta,\alpha)$-bicriteria approximation
algorithm for SCSC, and a similar result holds for the opposite direction.

Thus, the approximation results for SM with an approximate oracle could potentially be
translated into an approximation guarantee for SCSC with an approximate oracle.
However, the results of \citeauthor{horel2016} are for the special case of SCSK where the cost
function is cardinality. To the best
of our knowledge, we are the first to study submodular optimization with
oracle error and  general monotone submodular cost functions. In addition,
the feasibility guarantee provided under the
\citeauthor{iyer2013} framework is roughly $f(A) > (1 - 1/e)\tau$, so only
a fraction of the threshold $\tau$: in our work, we
obtain the feasibility $f(A) \ge \tau - \epsilon$.

SM with an approximate oracle has been studied under different models for the surrogate
$F$ than an $\epsilon$-approximation \cite{hassidim2017,singer2018}. These models are suited to applications where
the approximate oracle is due to noise.

An approximate oracle to a submodular function
results in a non-submodular function. 
Other works optimizing non-submodular or weakly submodular
functions include \citet{Bian2017,Chen2017,Kuhnle2018a}.


%


\subsection{Definitions}
\label{section:definitions}
The definitions presented in this section are used throughout the paper.

\paragraph{Notation}
%
Given a function $g:2^S\to\mathbb{R}_{\geq 0}$,
define $g_{\tau}:2^S\to\mathbb{R}_{\geq 0}$ to be
$g_{\tau}(X) = \min\{\tau, g(X)\}$ for all $X\subseteq S$.
In addition, we shorten the notation for marginal gain to be
$\Delta g(X,x) = g(X\cup\{x\})-g(X)$.

Given an instance of SCSC with cost function $c$ and benefit function $f$,
we define $c_{min} = \min_{x \in S}c(x)$,
$c_{max} = \max_{x \in S}c(x)$, $\alpha = \max_{x \in S}f(\{x\})$, and $\rho$ to be the curvature
of $c$.

Suppose at the end of a run of Algorithm \ref{algorithm:greedy}
there were $k$ iterations of the while loop.
Then we let $A_i$ be $A$ at the end of iteration $i \in\{1,...,k\}$,
$A_0=\emptyset$,
$\mu = \min\{f_{\tau}(A_i)-f_{\tau}(A_{i-1}):i\in\{1,...,k\}\}$,
and $\beta = \min\{\Delta f_{\tau}(A_i, x): i\in\{0,...,k\}, x\in S,
\Delta f_{\tau}(A_i, x) > 0\}$.

\paragraph{Greedy algorithm}
Pseudocode for the greedy algorithm that we analyze in Section \ref{section:mainresults}
for SCSC with an $\epsilon$-approximate oracle is given in Algorithm \ref{algorithm:greedy}.
Notice that when choosing an element at each iteration, we compute the marginal gain
of $F_{\tau}$ and not $F$. Algorithm \ref{algorithm:greedy}
was analyzed by \citeauthor{wan2010} for SCSC when a value oracle to $f$ is given.

\begin{algorithm}[tb]
  \caption{\texttt{greedy}$(F, c, \tau)$}
  \label{algorithm:greedy}
  \begin{algorithmic}

    \STATE {\bfseries Input:}A value oracle to $F: 2^S \to \mathbb{R}_{\geq 0}$,
      a value oracle to $c: 2^S \to \mathbb{R}_{\geq 0}$,
      and $\tau$.\\

      $F_{\tau} = \min\{F, \tau\}$

      $A = \emptyset$\;

      \WHILE {$F(A) < \tau$}
      \STATE {
        $u = \text{argmax}_{x \in S\setminus A}\Delta F_{\tau}(A,x)/c(x)$\;

        $A = A \cup \{u\}$\;
      }
      \ENDWHILE

      \RETURN $A$
    \end{algorithmic}
  \end{algorithm}

\paragraph{$\epsilon$-Approximate}
A function $F:2^S\to\mathbb{R}_{\geq 0}$ is $\epsilon$-approximate to $f$ if
for all $X\subseteq S$, $|f(X)-F(X)| \leq \epsilon$. Notice that we use
$\epsilon$-approximate in an absolute sense, in contrast to $\epsilon$-approximate in
a relative sense: for all $X\subseteq S$, $|f(X)-F(X)| \leq \epsilon f(X)$. The latter is
particularly useful if we are uncertain what range $f$ takes on, in which case it is difficult to make
meaningful requirements for additive noise.

Approximation in the relative sense can be converted into
approximation in the absolute sense. Suppose $F$ is an $\epsilon$-approximation to
$f$ in the relative sense. If $B$ is an upper bound on $f$, then $F$ is an $\epsilon B$-approximation to $f$ in
the absolute sense. Over the duration of Algorithm \ref{algorithm:greedy}, we can assume without loss
of generality that $\tau$ is an upper bound on $f$.

\paragraph{Curvature}
The approximation guarantees presented in our work will use the curvature of the cost function $c$,
as has been previously done for SCSC \cite{wan2010,soma2015}.
The curvature measures how modular\footnote{$c$
is modular if $c(X)=\sum_{x\in X}c(\{x\})$.} a function is.
The curvature $\rho$ of $c$ is defined as
$\rho = \max_{X \subseteq S}\sum_{x\in X}c(x)/c(X)$.
If $c$ is modular, $\rho=1$, otherwise $\rho > 1$ (since $c$ is submodular).

\paragraph{Bicriteria Approximation Algorithm}
We show in Section \ref{section:mainresults}
that Algorithm \ref{algorithm:greedy} is a bicriteria approximation algorithm
to SCSC, under certain conditions.
A bicriteria approximation algorithm approximates the feasibility constraint
($f(A) \geq \tau$) in addition to the objective (minimize $c$). In our case, the
feasibility guarantee is $f(A)\geq \tau - \epsilon$ if we have an
$\epsilon$-approximate oracle.

\section{Approximation Results}
\label{section:mainresults}
In this section, we analyze the approximation guarantee of the greedy algorithm
(Algorithm \ref{algorithm:greedy}) for SCSC with
an $\epsilon$-approximate oracle. Definitions used in this section can be found in
Section \ref{section:definitions}.

We first give a
formal statement and discussion of our ratios in Section \ref{section:ratios}.
In Section \ref{section:incomparable}, we prove that the two ratios presented in Section \ref{section:ratios}
are incomparable.
A sketch of the proofs of our results is presented in Section \ref{section:ratioproofs}.
Parts of the proofs not included in Section \ref{section:ratioproofs}
are in Appendices \ref{appendix:theorem1} and \ref{appendix:theorem2}.
The approximation guarantee of the greedy algorithm for SCSC with an
$\epsilon$-approximate oracle for the special case where the oracle is monotone submodular is in Appendix \ref{section:submodularF}.

\subsection{Approximation Guarantees of the Greedy Algorithm for SCSC with an $\epsilon$-Approximate Oracle}
\label{section:ratios}
We present two approximation guarantees of the greedy algorithm (Algorithm \ref{algorithm:greedy}) for SCSC
with an $\epsilon$-approximate oracle in Theorem \ref{ratio} and Theorem \ref{ratio2}.
The guarantee in Theorem \ref{ratio} corresponds more closely to
existing approximation guarantees of SCSC \cite{wolsey1982,wan2010,soma2015}
than that of Theorem \ref{ratio2}. However, in some cases, Theorem \ref{ratio2}
is easier to bound above. In general, the two ratios are incomparable;
that is, there exist instances of SCSC where each dominates the other,
as shown in Section \ref{section:incomparable}.

\begin{theorem}
\label{ratio}
Suppose we have an instance of SCSC with optimal solution $A^*$.
Let $F$ be a function that is $\epsilon$-approximate to $f$.

Suppose we run
Algorithm \ref{algorithm:greedy} with input $F$, $c$, and $\tau$.
Then $f(A) \geq \tau - \epsilon$.
And if $\mu > 4\epsilon c_{max}\rho/c_{min}$,
\begin{align*}
  c(A) \leq& \frac{\rho}{1-\frac{4\epsilon c_{max} \rho}{c_{min} \mu}}
             \left(\ln\left(\frac{\alpha}{\beta}\right) + 2\right)c(A^*).
\end{align*}
\end{theorem}

\paragraph{Discussion of Theorem \ref{ratio}}
In order for the ratio of Theorem \ref{ratio} to hold, $\epsilon$ must be small enough
relative to $\mu$ so that $\mu > 4\epsilon c_{max}\rho/c_{min}$.
The lower bound on $\mu$
is used to upper bound the error introduced by choosing elements with $F$ instead of $f$
in the proof of Theorem \ref{ratio} (see Section \ref{section:ratioproofs}).
Alternatively, we may ensure the approximation ratio of Theorem \ref{ratio} as long as
$\epsilon$ is sufficiently small by exiting
Algorithm \ref{algorithm:greedy} if $F_{\tau}(A_{i+1})-F_{\tau}(A_i)$ falls
below an input value. The feasibility guarantee is weakened since Algorithm \ref{algorithm:greedy}
does not necessarily run to completion, but not by much if $\epsilon$ is sufficiently small.
The details of this alternative approximation guarantee
can be found in Appendix \ref{section:submodularF}.

If $\epsilon=0$, i.e. we have an oracle to the benefit function $f$, then
the lower bound on $\mu$ is always satisfied and
the approximation ratio in Theorem \ref{ratio} nearly reduces to the ratio of
previous approximation ratios for SCSC \cite{wolsey1982,wan2010,soma2015}.
In particular, the approximation ratio reduces to
$\rho(\ln(\alpha/\beta) + 2)$. Compare this to the approximation ratio of
\citeauthor{soma2015}: $\rho(1 + 3\delta)(\ln(\alpha/\beta)+1)$ where $\delta$
is an input that is greater than 0.

Computing $\alpha, \beta$ and $\mu$ in Theorem \ref{ratio} requires evaluation of $f$.
It is therefore of interest whether an upper bound can be computed on the ratio
in Theorem \ref{ratio} for a given instance of SCSC and a solution provided by the greedy algorithm,
considering that we only have an oracle to $F$.
We assume that the curvature $\rho$ of $c$ can be computed
and focus on the values related to
$f$. Without an oracle to $f$, the
value $\alpha$ in Theorem \ref{ratio} cannot be computed
exactly, but $\alpha$ can be bounded above by using the oracle to $F$.
Similarly, $\mu$ and $\beta$ must be bounded below by using the oracle to $F$.
However, a positive lower bound on $\beta$ is problematic since it can be especially small
and fall below the oracle error. This motivates our second approximation ratio,
Theorem \ref{ratio2}.

\begin{theorem}
\label{ratio2}
Suppose we have an instance of SCSC with optimal solution $A^*$.
Let $F$ be a function that is $\epsilon$-approximate to $f$.

Suppose we run
Algorithm \ref{algorithm:greedy} with input $F$, $c$, and $\tau$.
Then $f(A) \geq \tau - \epsilon$.
And if $\mu > 4\epsilon c_{max}\rho/c_{min}$, then for any
$\gamma \in \left(0,1-4\epsilon c_{max}\rho/c_{min}\mu\right)$,
\begin{align*}
  c(A) \leq& \frac{\rho}{1-\frac{4\epsilon c_{max} \rho }{c_{min} \mu} - \gamma}
             \left(\ln\left(\frac{n\alpha\rho}{\gamma\mu}\right) + 2\right)c(A^*).
\end{align*}
\end{theorem}

\paragraph{Discussion of Theorem \ref{ratio2}}
If $\epsilon=0$, i.e. we have an oracle to the benefit function $f$, then
the lower bound on $\mu$ is always satisfied and
the approximation ratio in Theorem \ref{ratio2} is a new approximation
ratio for SCSC that is incomparable to those existing \cite{wolsey1982,soma2015}.

In contrast to the approximation guarantee of Theorem \ref{ratio}, the instance-dependent
$\beta$ has been replaced by
$\gamma$ in the approximation guarantee of Theorem \ref{ratio2}. Since we no longer need
a positive lower bound on $\beta$, this ratio was easy
to bound in Section \ref{section:experiments} by using the oracle to $F$.
Also, since $\beta$ is related to
the minimum marginal gain on an instance, in some sense this ratio is more robust
to the presence of very small marginal gains.

\subsection{Incomparability of Guarantees of Theorem \ref{ratio} and Theorem \ref{ratio2}}
\label{section:incomparable}
In this section, we give examples that show that the approximation guarantees of Theorem
\ref{ratio} and of Theorem \ref{ratio2} are incomparable; for each guarantee,
there exists an instance of SCSC where that guarantee is better than the other.

\paragraph{Examples}
We consider an instance of the Influence Threshold (IT) problem as defined in Appendix \ref{appendix:influence}
under the independent cascade model of influence \cite{kempe2003}.

Let $n\geq 3$.
We construct a graph where $n-2$ vertices are in a clique, and all edge weights
within the clique have weight 1. One remaining vertex is connected
to the clique by an edge of weight $\sigma \in (0,1)$, and the other has degree 0.

Suppose we have an instance of SCSC where $c$ is cardinality and $\tau=n-1+\sigma$.
If we run Algorithm \ref{algorithm:greedy} with input $f,c,\tau$, Algorithm
\ref{algorithm:greedy} will return a single vertex from the clique and then the
vertex with degree 0. In addition,
$\alpha=n-2+\sigma$, $\mu=1$, and $\beta=1-\sigma$.
Therefore the ratio from Theorem \ref{ratio} is
\begin{align}
  \ln\left(\frac{n-2+\sigma}{1-\sigma}\right)+2 \label{ratio:1-ex}
\end{align}
and the ratio from Theorem \ref{ratio2} is for any $\gamma\in(0,1)$
\begin{align}
  \frac{1}{1-\gamma}\ln\left(\frac{n(n-2+\sigma)}{\gamma}\right)+2 \label{ratio:2-ex}
\end{align}
If we choose $\sigma$ sufficiently close to 1,
ratio (\ref{ratio:1-ex}) gets arbitrarily large; hence,
there exists some $\gamma$ where ratio (\ref{ratio:2-ex}) is smaller than ratio
(\ref{ratio:1-ex}).
On the other hand, as $\sigma$ approaches 0, ratio (\ref{ratio:1-ex}) approaches $\ln (n - 2) + 2$.
However, for any $\gamma \in (0,1)$, ratio (\ref{ratio:2-ex}) is at least $\ln (n(n-2)) + 2$.

\subsection{Proof Sketches of Theorem \ref{ratio} and \ref{ratio2}}
\label{section:ratioproofs}

In this section, we present a sketch of the proofs of Theorem \ref{ratio} and
Theorem \ref{ratio2}. The full proof for Theorem \ref{ratio} and for Theorem \ref{ratio2}
can be found in Appendix \ref{appendix:theorem1} and Appendix \ref{appendix:theorem2}
respectively.
Recall that definitions can be found in Section \ref{section:definitions}
and notation in Section \ref{section:ratios}.

\paragraph{Proof Sketch of Theorem \ref{ratio}}
\setcounter{equation}{0}

The feasibility guarantee is clear from the stopping condition on the greedy algorithm:
$f(A) \geq F(A) - \epsilon \geq \tau - \epsilon$.

We now prove the upper bound on $c(A)$ if $\mu > 4\epsilon c_{max}\rho/c_{min}$.
Without loss of generality we re-define $f=\min\{f,\tau\}$ and $F=\min\{F,\tau\}$.
This way, $f=f_{\tau}$ and $F=F_{\tau}$.
Notice that this does not change that $F$ is an $\epsilon$-approximation of $f$.

Let $x_1,...,x_k$ be the elements of $A$ in the order that they were chosen
by Algorithm \ref{algorithm:greedy}. If $k=0$, then $c(A)=0$ and the approximation ratio is clear.
For the rest of the proof, we assume that $k\geq 1$.

We define a sequence of elements $\tilde{x}_1,...,\tilde{x}_k$ where
\begin{align*}
  \tilde{x}_i =
    \text{argmax}_{x \in S\setminus A_{i-1}}\frac{\Delta f(A_{i-1}, x)}{c(x)}.
\end{align*}
$\tilde{x}_i$ has the most cost-effective marginal gain of being added to $A_{i-1}$
according to $f$, while $x_i$ has the most cost-effective marginal gain of being added to
$A_{i-1}$ according to $F$.
Note that the same element can appear multiple times in the sequence
$\tilde{x}_1,...,\tilde{x}_k$. In addition, we have the following lower bound on $\Delta f(A_{i-1}, \tilde{x}_i)$:
\begin{align}
  \Delta f(A_{i-1}, \tilde{x}_i)
  \geq \frac{c(\tilde{x}_i)}{c(x_i)}\Delta f(A_{i-1}, x_i)
  \geq \frac{c_{min}}{c_{max}}\mu \label{peqn2}.
\end{align}

Our argument to bound $c(A)$ will follow
the following three steps:
\textbf{(a)} We bound $c(A)$ in terms of the costs of the elements
$\tilde{x}_1,...,\tilde{x}_k$.
\textbf{(b)} We charge the elements
of $A^*$ with the costs of the elements $\tilde{x}_1,...,\tilde{x}_k$,
and bound $c(A)$ in terms of the total charge on all elements in $A^*$.
\textbf{(c)} We bound the total charge on the elements of $A^*$ in terms of $c(A^*)$.

\paragraph{(a)} First, we bound $c(A)$ in terms of the costs of the elements
$\tilde{x}_1,...,\tilde{x}_k$.
At iteration $i$ of Algorithm \ref{algorithm:greedy}, the most cost-effective element
to add to $A_{i-1}$ according to $F$ is $x_i$.
Using the fact that $F$ is $\epsilon$-approximate to $f$, we can bound
how much more cost-effective $\tilde{x}_i$ is compared to $x_i$ according to $f$
as follows:
\begin{align}
  &\frac{c(\tilde{x}_i)}{\Delta f(A_{i-1}, \tilde{x}_i)} + \alpha_i
  \geq \frac{c(x_i)}{\Delta f(A_{i-1}, x_i)} \label{peqn4} \\
  &\text{where }\alpha_i = \frac{2\epsilon(c(x_i)+c(\tilde{x}_i))}{\Delta f(A_{i-1}, \tilde{x}_i)\Delta f(A_{i-1}, x_i)}.\nonumber
\end{align}

Inequality (\ref{peqn4}) and the submodularity of $c$ imply that
\begin{align}
 c(A) &\leq \sum_{i=1}^k c(x_i)
      = \sum_{i=1}^k \Delta f(A_{i-1},x_i)
         \frac{c(x_i)}{\Delta f(A_{i-1},x_i)} \nonumber \\
      &\leq \sum_{i=1}^k \Delta f(A_{i-1}, {x_i})
          \left(\frac{c(\tilde{x}_i)}{\Delta f(A_{i-1}, \tilde{x}_i)}
          + \alpha_i\right). \label{peqn5}
\end{align}

We now bound the second term on the right side of Equation (\ref{peqn5}) by
\begin{align*}
  \sum_{i=1}^k \Delta f(A_{i-1}, {x_i})\alpha_i
  &= \sum_{i=1}^k \frac{2\epsilon(c(\tilde{x}_i) + c(x_i))}{\Delta f(A_{i-1},\tilde{x}_i)} \\
  &\leq \sum_{i=1}^k \frac{2\epsilon c(x_i)}{\Delta f(A_{i-1},x_i)}
      + \frac{2\epsilon c(x_i)}{\Delta f(A_{i-1},\tilde{x}_i)} \\
  &\leq \frac{4\epsilon c_{max}}{c_{min}\mu}\sum_{i=1}^k c(x_i)
  \leq \frac{4\epsilon c_{max}\rho}{c_{min}\mu}c(A).
\end{align*}

Applying this bound to (\ref{peqn5}) gives us the following bound on $c(A)$
in terms of the costs of the elements $\tilde{x}_1,...,\tilde{x}_k$:
\begin{align}
  \left( 1 - \frac{4\epsilon c_{max}\rho}{c_{min}\mu} \right)c(A)
  \leq \sum_{i=1}^k \frac{\Delta f(A_{i-1},x_i)}{\Delta f(A_{i-1},\tilde{x}_i)} c(\tilde{x}_i). \label{peqn7}
\end{align}

\paragraph{(b)} Next, we charge the elements
of $A^*$ with the costs of the elements $\tilde{x}_1,...,\tilde{x}_k$,
and bound $c(A)$ in terms of the total charge on all elements in $A^*$.
By this we mean that we give each $y\in A^*$ a portion of the total cost of
the elements $\tilde{x}_1,...,\tilde{x}_k$. In particular,
we give each $y\in A^*$ a charge of $w(y)$, defined by
\begin{align*}
  w(y) &= \sum_{i=1}^k(\pi_i(y) - \pi_{i+1}(y))\omega_i
  \text{, where }
  \omega_i = \frac{c(\tilde{x}_i)}{\Delta f(A_{i-1}, \tilde{x}_i)}, \\
  &\text{and }
  \pi_i(y) = \begin{cases}
        \Delta f(A_{i - 1}, y) & i \in \{1,...,k\} \\
        \Delta f(A, y) & i = k+1
     \end{cases}.
\end{align*}
Recall that $\Delta f(A_{i-1}, \tilde{x}_i) > 0$ for all $i$ by Equation (\ref{peqn2}),
and so we can define $\omega_i$ as above. \citeauthor{wan2010} charged the elements of $A^*$ with the
cost of elements $x_1,...,x_k$ analogously to the above. We charge with the cost of elements
$\tilde{x}_1,...,\tilde{x}_k$ because they exhibit diminishing cost-effectiveness, i.e.
$\omega_i-\omega_{i-1}\geq 0$ for all $i\in\{1,...,k\}$, which is needed to proceed with the argument.
Because we choose $x_1,...,x_k$ with
$F$, which is not monotone submodular, $x_1,...,x_k$ do not exhibit diminishing cost-effectiveness
even if we replace $f$ with $F$ in the definition of $w(y)$ above.

Using Equation (\ref{peqn7}) and an argument similar to \citeauthor{wan2010}, we may work out that
\begin{align}
  \left(1 - \frac{4\epsilon c_{max}\rho}{\mu c_{min}}\right)c(A)
  \leq \sum_{y\in A^*}w(y) + \Delta f(A,y)\omega_k.\label{peqnn1}
\end{align}
$f(A)$ is not necessarily $\tau$ since the stopping condition for Algorithm \ref{algorithm:greedy}
is only that $F(A)\geq\tau$. In this case, if $\Delta f(A,y)\neq 0$ for $y\in A^*$ (which implies that
$y\notin A_{k-1}$)
then by the submodularity of $f$
\begin{align*}
  \Delta f(A,y)\omega_k
  &\leq \Delta f(A,y)\frac{c(y)}{\Delta f(A_{k-1},y)}
  \leq c(y).
\end{align*}
Therefore we can bound $c(A)$ in terms of the total charge on all
elements in $A^*$:
\begin{align}
  \left(1 - \frac{4\epsilon c_{max}\rho}{\mu c_{min}}\right)c(A)
  \leq \sum_{y\in A^*}w(y) + \rho c(A^*). \label{peqn11}
\end{align}

\paragraph{(c)} Now, we bound the total charge on the elements of $A^*$ in terms of $c(A^*)$.

We first define a value $\ell_y$ for every $y\in A^*$. For each $y\in A^*$,
if $\pi_1(y) = 0$ we set $\ell_y=0$,
otherwise $\ell_y$ is the value in $\{1,...,k\}$ such that if $i\in \{1,...,\ell_y\}$
then $\pi_i(y) > 0$, and if $i\in \{\ell_y+1,...,k\}$ then $\pi_i(y) = 0$. Such an $\ell_y$
can be set since $f$ is monotone submodular.
Then
\begin{align}
  \sum_{y\in A^*}w(y) &= \sum_{y\in A^*}\sum_{i=1}^{\ell_y}(\pi_i(y) - \pi_{i+1}(y))\omega_i \nonumber \\
       &= \sum_{y\in A^*}c(y)\left( \ln(\pi_1(y))-\ln(\pi_{\ell_y}(y)) + 1\right) \nonumber \\
       &\leq \rho\left(\ln\left(\frac{\alpha}{\beta}\right) + 1\right)c(A^*). \label{rpeqn13}
\end{align}

Finally, we combine inequality (\ref{peqn11}) and inequality (\ref{rpeqn13}) to see that
\begin{align*}
  \left(1 - \frac{4\epsilon c_{max}\rho}{c_{min}\mu}\right)c(A)
  &\leq \rho\left(\ln\left(\frac{\alpha}{\beta}\right)+2\right)c(A^*).
\end{align*}
If $\mu > (4\epsilon c_{max} \rho)/(c_{min})$, this completes the proof of the approximation
guarantee in the theorem statement. \qed

\paragraph{Proof Sketch of Theorem \ref{ratio2}}
\setcounter{equation}{0}
The argument for the proof of Theorem \ref{ratio2} is the same as Theorem \ref{ratio},
except for part \textbf{(c)}.
In particular, we have gotten to the point of the proof of Theorem \ref{ratio} where we have proven that
\begin{align}
  \left(1-\frac{4\epsilon c_{max}\rho}{c_{min}\mu}\right)c(A) \leq
    \sum_{y\in A^*}w(y) + \rho c(A^*). \label{peqn13}
\end{align}

Let $\lambda > 0$. Then we define a value $m_y$ for every $y\in A^*$ that is similar to
$\ell_y$ but for when $\pi_i(y)$ falls below $\lambda$: For each $y\in A^*$,
if $\pi_1(y) \leq \lambda$ we set $m_y=0$,
otherwise $m_y$ is the value in $\{1,...,k\}$ such that if $i\in\{1,...,m_y\}$ then
$\pi_i(y) > \lambda$, and if $i\in\{m_y+1,...,k\}$ then $\pi_i(y)\leq\lambda$.
Such an $m_y$ can be set since $f$ is monotone submodular. We may then use a similar
analysis as in the proof of Theorem \ref{ratio} to see that
\begin{align*}
  w(y) &\leq c(y)\left(\ln\left(\frac{\alpha}{\lambda}\right) + 1\right)
  + \sum_{i=m_y+1}^{k}(\pi_i(y)-\pi_{i+1}(y))\omega_i \\
  &\leq c(y)\left(\ln\left(\frac{\alpha}{\lambda}\right) + 1\right) +\frac{\lambda\rho}{\mu}c(A).
\end{align*}
By combining Equations (\ref{peqn13}) and the bound on $w(y)$, we have that
\begin{align*}
  \left(1 - \frac{4\epsilon c_{max}\rho}{c_{min}\mu} - \frac{\lambda\rho n}{\mu}\right)c(A)
      &\leq \rho\left(\ln\left(\frac{\alpha}{\lambda}\right)+2\right)c(A^*).
\end{align*}

If we set $\lambda = (\gamma\mu)/(n\rho)$ we have the approximation ratio in
the theorem statement.
\qed

\section{Application and Experiments}
\label{section:experiments}
In this section, we compute the approximation ratios stated in Theorems \ref{ratio}
and \ref{ratio2} on instances of the Influence Threshold problem (IT), a special case
of SCSC. We use the non-submodular approximate reachability oracle that has been
proposed by \citeauthor{cohen2014} (\citeyear{cohen2014}).

\paragraph{Influence Threshold Problem (IT)}
Let $G=(V,E)$ be a social network where vertices $V$ represents users and
edges $E$ represent social connections.
Suppose that $G_1=(V,E_1),...,G_N=(V,E_N)$, where $E_i\subseteq E$, represent
$N$ instances of ``alive" social connections.
In an instance, activation of users in the social network starts from an
initial seed set and then propagates across edges.
$c:2^V\to\mathbb{R}_{\geq 0}$ is a monotone submodular function that gives the cost of
seeding a set of users. For $X\subseteq V$,
$f(X)$ is the average number of reachable vertices from $X$ over the $N$ instances.
Given threshold $\tau \leq f(V)$, the Influence Threshold (IT) problem is to
find the seed set $\text{argmin}\{c(X)| X\subseteq V, f(X)\geq\tau\}$.

Our definition of IT follows the simulation-based model of influence, as opposed to directly using a model
such as Independent Cascade (IC) and defining $f$ as the expected number of influenced vertices
\cite{kempe2003}.
The IC model is commonly approximated by
the simulation-based model, since computing the expected influence under the IC model
is $\#P$-hard \cite{chen2010}. In addition, the simulation-based model approximates the IC model arbitrarily well
by choosing sufficiently large $N$.
For more details on approximation of the IC model by the simulation-based model, see Appendix \ref{appendix:influence}.

Variations of IT where $c$ is cardinality \cite{he2014,dinh2014,kuhnle2017} or modular \cite{goyal2013,han2017}
have been studied in the influence literature.
Notice the difference between IT and the Influence Maximization (IM) problem\footnote{
Given a budget $\kappa$, IM is to determine the set $A$ such that $c(A)\leq\kappa$ and $f(A)$
is maximized.} \cite{kempe2003,li2017approximate}.
To the best of our knowledge,
our approximation results are the first for IT with a general monotone submodular cost function.

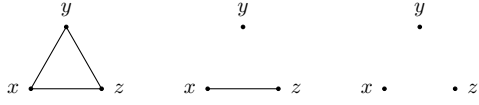
\begin{figure}
  \label{fig:realize}
  \centering
\begin{tikzpicture}[scale=0.47, every node/.style={scale=0.8}]
  \draw [fill] (0,0) circle [radius=0.05];
  \node at (-0.5,0) {$x$};
  \draw [fill] (1,1.75) circle [radius=0.05];
  \node at (1,2.25) {$y$};
  \draw [fill] (2,0) circle [radius=0.05];
  \node at (2.5,0) {$z$};

  \draw (0,0) -- (1,1.75);
  \draw (1,1.75) -- (2,0);
  \draw (0,0) -- (2,0);

  \draw [fill] (5,0) circle [radius=0.05];
  \node at (4.5,0) {$x$};
  \draw [fill] (6,1.75) circle [radius=0.05];
  \node at (6,2.25) {$y$};
  \draw [fill] (7,0) circle [radius=0.05];
  \node at (7.5,0) {$z$};

  \draw (5,0) -- (7,0);

  \draw [fill] (10,0) circle [radius=0.05];
  \node at (9.5,0) {$x$};
  \draw [fill] (11,1.75) circle [radius=0.05];
  \node at (11,2.25) {$y$};
  \draw [fill] (12,0) circle [radius=0.05];
  \node at (12.5,0) {$z$};

\end{tikzpicture}
\caption{Instance $G_1, G_2$ and $G_3$ (in that order from left to right) from the proof of
  Proposition \ref{prop:cohen}.}
\end{figure}

\begin{figure}
  \label{fig:mapping}
  \centering
\begin{tikzpicture}[scale=0.47, every node/.style={scale=0.8}]

\draw (0,0) -- (17.25,0);

\node at (0,0.75) {$0$};
\draw (0,-0.2) -- (0,0.2);

\node at (0.6,0.75) {$r_z^2$};
\draw (0.6,-0.2) -- (0.6,0.2);

\node at (1.7,0.75) {$r_x^2$};
\draw (1.7,-0.2) -- (1.7,0.2);

\node at (3.5,0.75) {$r_z^1$};
\draw (3.5,-0.2) -- (3.5,0.2);

\node at (4.5,0.75) {$r_z^3$};
\draw (4.5,-0.2) -- (4.5,0.2);

\node at (6,0.75) {$r_y^3$};
\draw (6,-0.2) -- (6,0.2);

\node at (12,0.75) {$r_y^2$};
\draw (12,-0.2) -- (12,0.2);

\node at (13.2,0.75) {$r_y^1$};;
\draw (13.2,-0.2) -- (13.2,0.2);

\node at (14,0.75) {$r_x^1$};
\draw (14,-0.2) -- (14,0.2);

\node at (16,0.75) {$r_x^3$};
\draw (16,-0.2) -- (16,0.2);

\node at (17.25,0.75) {$1$};
\draw (17.25,-0.2) -- (17.25,0.2);

\end{tikzpicture}
\caption{The mapping for each vertex, instance pair to the interval $[0,1]$ for the proof of Proposition \ref{prop:cohen}.
  $r_u^i$ is the mapped value of vertex $u$, instance $G_i$.}
\end{figure}
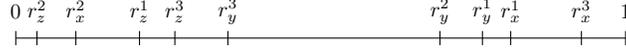

\paragraph{The Approximate Average Reachability Oracle of \citeauthor{cohen2014}}
\begin{figure*}
  \centering
  \hspace{-10pt}
  \subfigure[][facebook, r1] {
    \label{fig1}
    \includegraphics[width=0.25\textwidth]{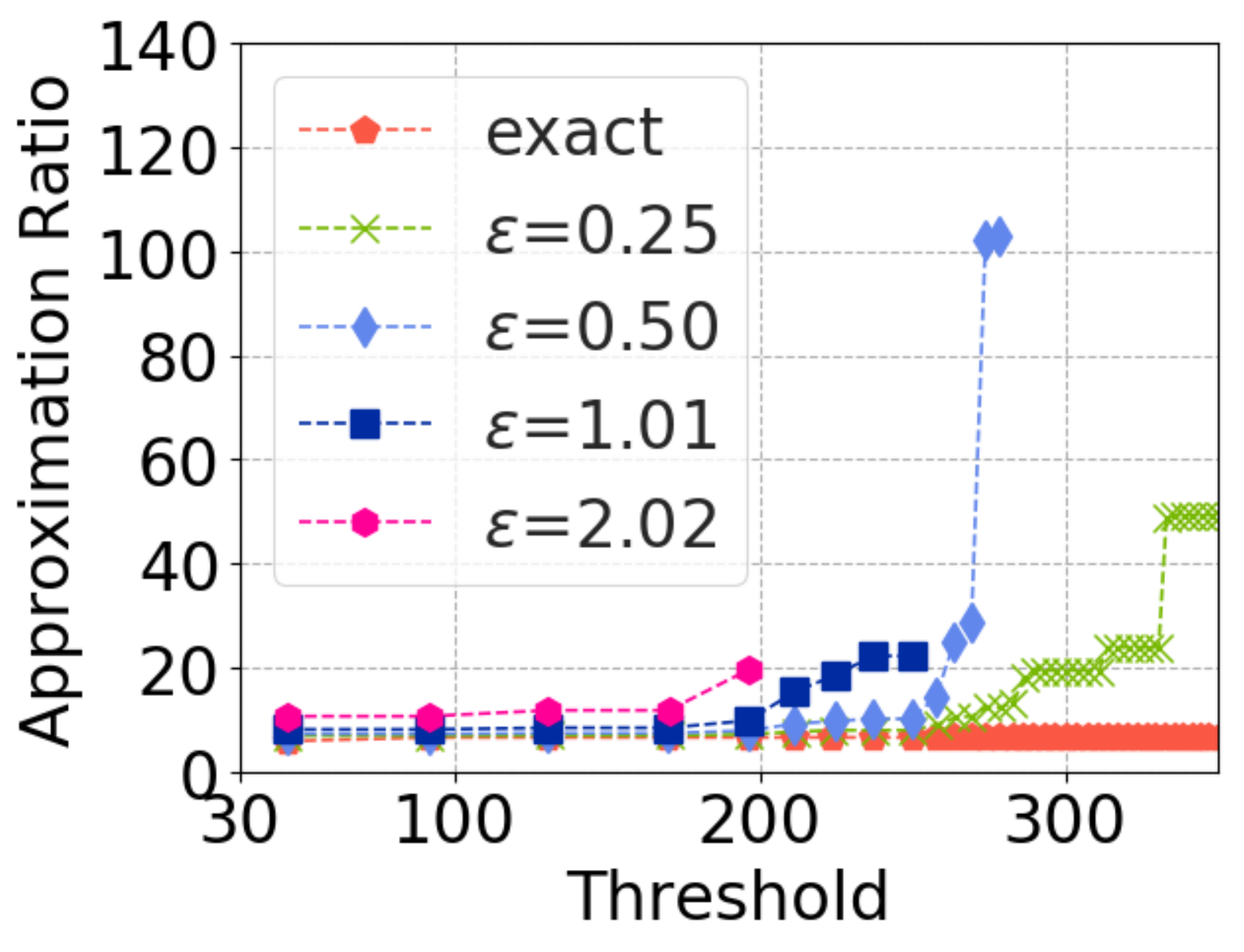}
  }
  \hspace{-10pt}
  \subfigure[][facebook, r2] {
    \label{fig2}
    \includegraphics[width=0.25\textwidth]{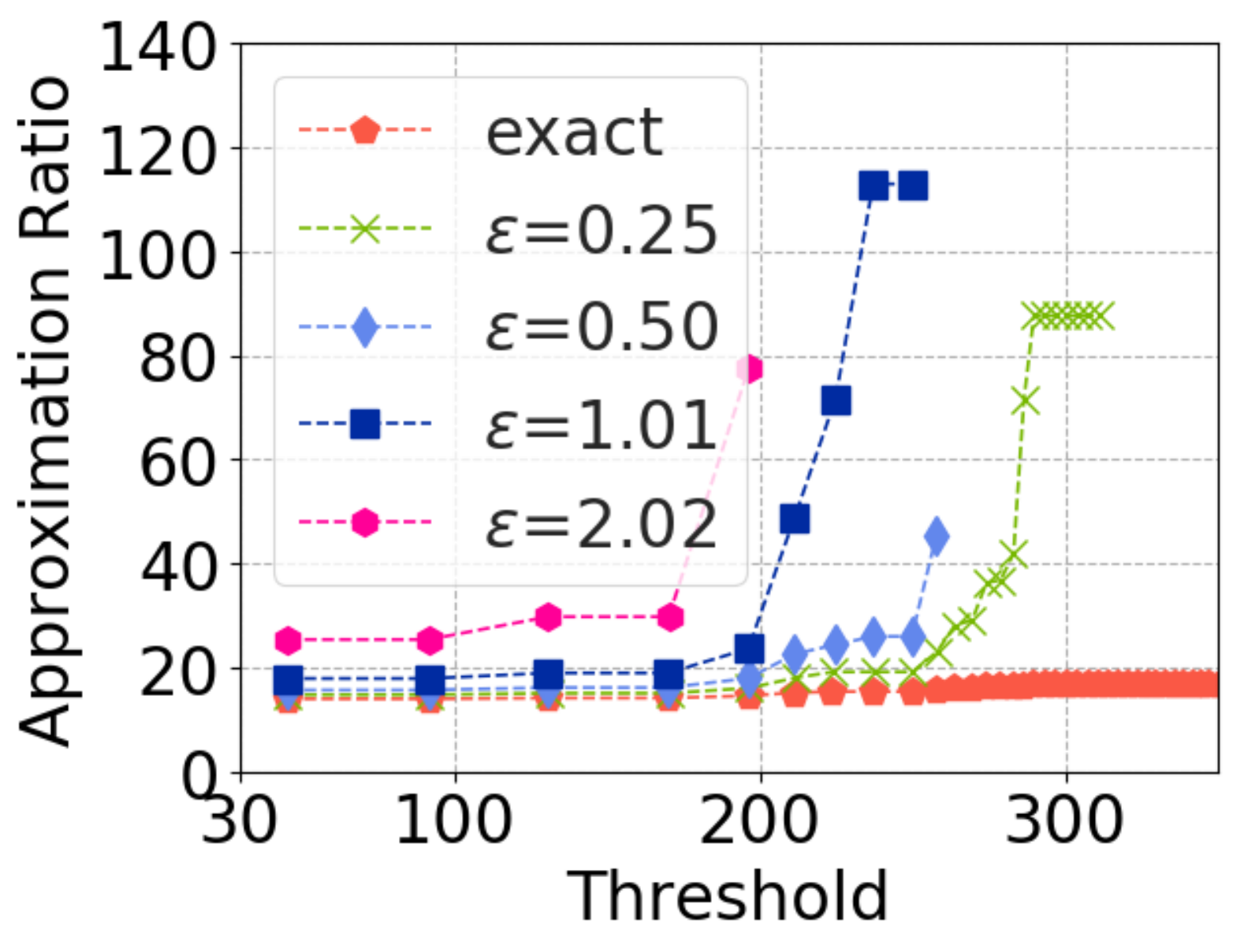}
  }
  \hspace{-10pt}
  \subfigure[][GrQc, r1] {
    \label{fig3}
    \includegraphics[width=0.25\textwidth]{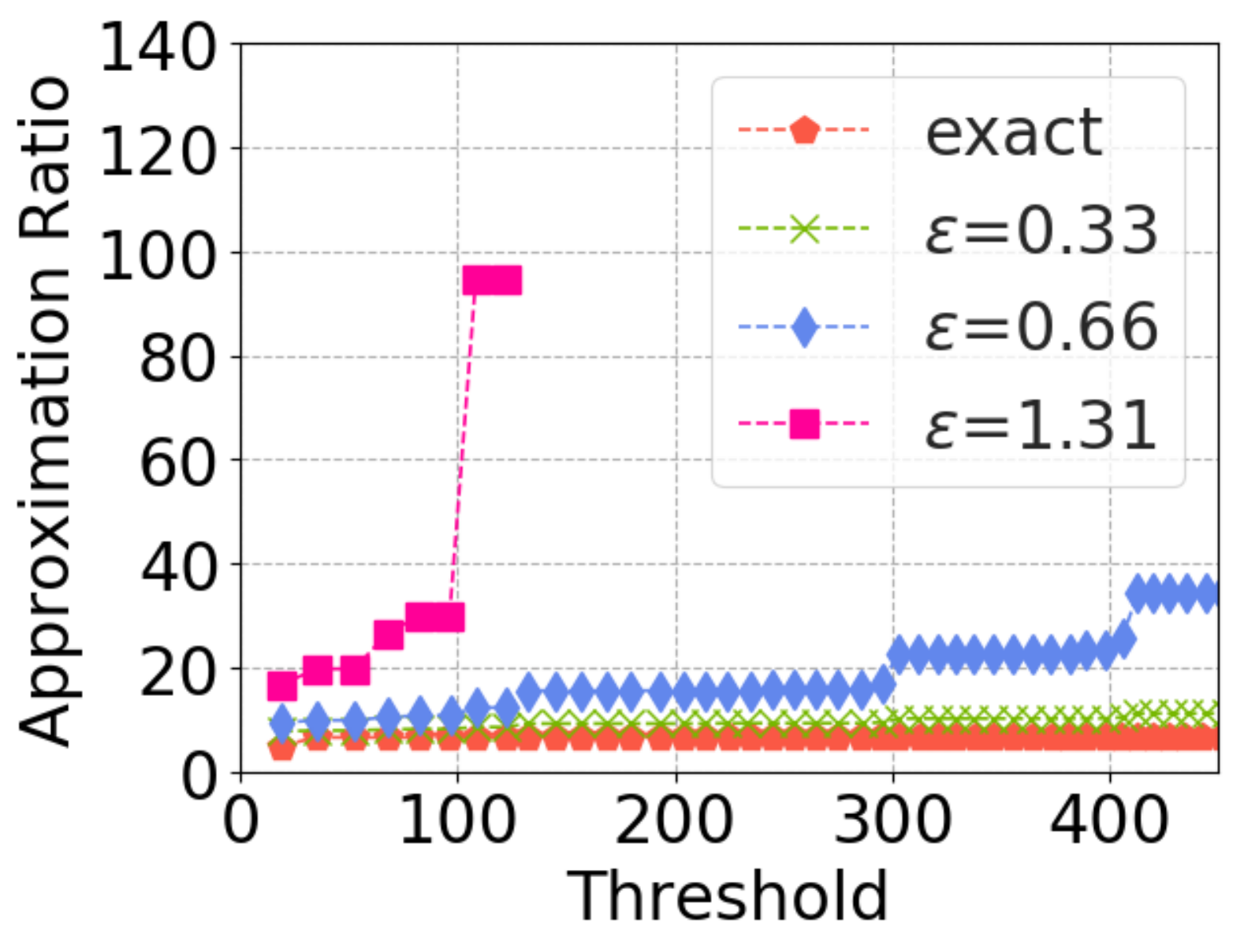}
  }
  \hspace{-10pt}
  \subfigure[][GrQc, r2] {
    \label{fig4}
    \includegraphics[width=0.25\textwidth]{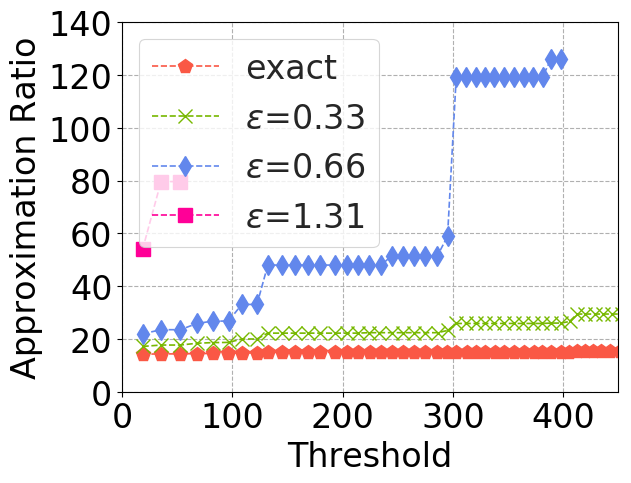}
  }
  \caption{The approximation ratios of Theorem \ref{ratio} (r1) and an upper bound on that of Theorem \ref{ratio2} (r2)
  at thresholds indicated by the markers.}
\end{figure*}


In order to have value oracle access to $f$ in the IT problem, the instances $G_1,...,G_N$ must be
stored. In addition, to compute $f(X)$ for $X\subseteq V$ the reachable vertices from $X$ must be computed
for each of the $N$ instances. If $N$ is large, this is not scalable to large influence instances \cite{cohen2014}.

Motivated by this, \citeauthor{cohen2014} (\citeyear{cohen2014}) proposed using an approximate
average reachability oracle in place of $f$ that is based on bottom-$k$ min-hash sketches \cite{cohen1997}.
Given $k\in\mathbb{Z}_{>0}$, the approximate average reachability oracle $F$ is constructed as follows:
For every vertex, instance pair $(v,i)\in V\times \{1,...,N\}$ a random rank
value $r_v^i$ is drawn from the uniform distribution on $[0,1]$.
For every vertex $u \in V$, the combined reachability sketch $X_u$ of $u$ is the smallest
$k$ values from the set $\{r_v^i: v \text{ is reachable from } u \text{ on instance } G_i \}$.
$X_u$ is stored for all $u\in V$.
Note that generating $X_u$ for all $u\in V$ does not require all the instances $G_1,...,G_N$ to be
stored at the same time.

Let $X\subseteq S$. If $|\cup_{u\in X}X_u|< k$, then $F(X)=|\cup_{u\in X}X_u|/N$.
Otherwise, let $t$ be the $k$-th
smallest value in $\cup_{u\in X}X_u$. Then $F(X) = (k-1)/(Nt)$.

$F$ can be made an $\epsilon$-approximation to $f$ by choosing sufficiently large $k$:
For $c>2$, if $k=c\epsilon^{-2}\log(n)$ then the relative error of all queries over
the duration of the greedy algorithm is within $\epsilon$ with probability at least $1-1/n^{c-2}$
\cite{cohen2014}. The relative error can be converted to absolute error as described in Section
\ref{section:definitions}.

\begin{proposition}
  \label{prop:cohen}
  The approximate average reachability oracle of \citeauthor{cohen2014} is non-submodular.
\end{proposition}
\begin{proof}
Consider an instance of IT where $V = \{x,y,z\}$ and three instances $G_1,G_2,G_3$ are as
depicted in Figure 1.

Suppose that we construct the approximate reachability oracle of \citeauthor{cohen2014} with
$k=5$, and the randomly generated mapping from vertex, instance pairs to the interval $[0,1]$
is as depicted in Figure 2. Then $X_x=\{r_z^2,r_x^2,r_z^1,r_y^1,r_x^1\}$,
$X_y=\{r_z^1,r_y^3,r_y^2,r_y^1,r_x^1\}$, and $X_z=\{r_z^2,r_x^2,r_z^1,r_z^3,r_y^1\}$.

Consider adding $z$ to the sets $A=\{x\}$ and $B=\{x,y\}$. Then
\begin{align*}
  \Delta F(A,z) &= (k-1)/(3r_y^1)-(k-1)/(3r_x^1), \\
  \Delta F(B,z) &= (k-1)/(3r_y^3)-(k-1)/(3r_y^2).
\end{align*}
If $r_y^1 \approx r_x^1$, but $r_y^3$ is sufficiently smaller than $r_y^2$,
$\Delta F(A,z) < \Delta F(B,z)$ despite $A\subseteq B$.
\end{proof}

The effectiveness of the approximate reachability oracle of \citeauthor{cohen2014} has
been extensively evaluated for both IM \cite{cohen2014} as well as
IT \cite{kuhnle2017}, both with uniform cost. In particular, the
approximate reachability oracle of \citeauthor{cohen2014} was demonstrated to be
significantly faster than alternative approaches such as TIM \cite{tang2014}.

\paragraph{Experimental Setup}
We use two real social networks: the Facebook ego network \cite{leskovec2012},
and the ArXiV General Relativity collaboration
network \cite{leskovec2007}, which we refer to as GrQc.
Influence propagation follows the Independent Cascade (IC) model \cite{kempe2003}.
The average reachability oracle, $f$, is over random realizations of the influence graph.
The approximate average reachability oracle of \citeauthor{cohen2014}, $F$, is computed over these
realizations with various oracle errors $\epsilon$ and the greedy algorithm is run using these oracles.
For comparison, we also run the greedy algorithm with $f$.

For the cost function, we
choose a cost $c_v$ for each $v\in V$ by sampling from a normal distribution
and then define $c(X)=\sum_{x\in X}c_x$. We include additional experiments
with non-modular cost functions in Appendix \ref{appendix:influence}.
Additional details about the experimental setup can be found in Appendix \ref{appendix:influence}.

\paragraph{Approximation Ratio of the Greedy Algorithm}
As we described in Section \ref{section:ratios}, if the minimum marginal gain of any element
added to the greedy solution is sufficiently large then
Theorems \ref{ratio} and \ref{ratio2} are approximation ratios.

For each run of the greedy algorithm with the approximate oracle $F$
where the minimum marginal gain was sufficiently large, we compute the ratio of
Theorem \ref{ratio} exactly by querying $f$ (``r1'', Figures \ref{fig1} and \ref{fig3})
and we compute an upper bound on
the ratio of Theorem \ref{ratio2} with only $F$ using
the argument described in the discussion of Theorem \ref{ratio2}
(``r2'', Figures \ref{fig2} and \ref{fig4}). Details of how we selected $\gamma$
for the ratio of Theorem \ref{ratio2} can be found in Appendix \ref{appendix:influence}.
We were generally unable to compute an upper bound on the ratio presented in Theorem
\ref{ratio} with just an oracle to $F$ because the $\beta$ term was too small.

In addition, we ran the greedy algorithm for the exact oracle $f$ and computed the
ratios of Theorem \ref{ratio} and Theorem \ref{ratio2} exactly where $\epsilon=0$.
Note that the ratio of Theorem \ref{ratio} when $\epsilon=0$ reduces to the
approximation ratio for SCSC given by \citeauthor{soma2015} (\citeyear{soma2015}).

\paragraph{Results}
  The approximation ratios on the Facebook dataset
  are plotted in Figures \ref{fig1} and \ref{fig2}.
  A marker indicates the threshold $\tau$ given as input to the greedy algorithm,
  and the corresponding ratio for that run.
  We chose the $F$ values at each
  step of the greedy algorithm to be the thresholds.
  If the minimum marginal gain of a run was too small relative to $\epsilon$
  for a threshold, or the approximation was greater than 140, then no marker is
  plotted.

  As expected, smaller oracle error $\epsilon$
  implies better approximation ratios. For the same $\epsilon$, the ratio
  of Theorem \ref{ratio} is better than the upper bound on the ratio of Theorem \ref{ratio2}.
  But, recall that the latter can be computed only querying $F$.
  As the threshold increases, the ratio degrades
  because elements with smaller marginal gain are being added into the solution set, although the effect
  is smaller with smaller $\epsilon$.

  Qualitatively similar results are shown for the GrQc dataset in Fig. \ref{fig3} and \ref{fig4}.
  The ratios degrade less quickly in
  GrQc compared to Facebook: this is because in Facebook there are a few vertices that can be selected with
  high marginal gain, and then after that the marginal gains quickly decrease. In contrast, the marginal
  gains for GrQc are more sustained over the course of the greedy algorithm.

\clearpage
\section{Acknowledgements}
This work was supported in part by NSF CNS-1814614, NSF EFRI 1441231,
and DTRA HDTRA1-14-1-0055.
Victoria G. Crawford was supported by a Harris Corporation Fellowship.
We thank the anonymous reviewers for their helpful feedback.
\bibliography{arxiv,mend}
\bibliographystyle{abbrvnat}
\onecolumn
\appendix

\section{Additional Approximation Results}
\label{section:submodularF}

\subsection{The case of monotone submodular $F$}
The main difficulty in proving Theorems \ref{ratio} and \ref{ratio2} in Section
\ref{section:ratios} is that the $\epsilon$-approximate oracle $F$ is not monotone submodular.
In the event that $F$ is monotone submodular, existing results for SCSC \cite{wolsey1982,wan2010,soma2015}
can be translated into results for SCSC with an
$\epsilon$-approximate oracle, as the following proposition shows.
\begin{proposition}
  \label{translate}
  Let $\mathcal{A}$ be a bicriteria approximation algorithm for SCSC
  that takes as input $f,c,\tau$ and
  has approximation ratio $\alpha$, and feasibility guarantee $\tau - \delta$.

  Suppose that we have two instances of SCSC: instance $\mathcal{I}_1$ with $f,c,\tau$
  and instance $\mathcal{I}_2$ with $F,c,\tau-\epsilon$
  where $F$ is an $\epsilon$-approximation to $f$.
  Then if we run $\mathcal{A}$ for $\mathcal{I}_2$ and $\mathcal{A}$ returns the set
  $A$, it is guaranteed that $c(A) \leq \alpha c(A^*)$ and $f(A) \geq \tau-2\epsilon-\delta$
  where $A^*$ is the optimal solution to instance $\mathcal{I}_1$.
\end{proposition}
\begin{proof}
  $A^*$ is a feasible solution to $\mathcal{I}_2$ since
  $F(A^*) \geq f(A^*)-\epsilon \geq \tau-\epsilon.$
  Therefore $\alpha c(A^*)\geq c(A)$.
  For the feasibility, we have that
  $f(A) \geq F(A) - \epsilon \geq \tau -2\epsilon - \delta.$
\end{proof}

\subsection{An alternative version of Theorem \ref{ratio}}
The definitions and notation used in this section can be found in Section \ref{section:definitions} of the paper.

The version of Theorem \ref{ratio} in Section \ref{section:mainresults} requires that
$\epsilon$ be small relative to the minimum marginal gain of an element added to
the greedy set, $\mu$, over the duration of Algorithm \ref{algorithm:greedy}.
In particular, $\mu > 4\epsilon c_{max}\rho/c_{min}$.
Alternatively, we may ensure the approximation ratio of Theorem \ref{ratio} if $\epsilon$ is sufficiently
small by exiting Algorithm \ref{algorithm:greedy} if $F_{\tau}(A_i)-F_{\tau}(A_{i-1})$ falls
below an input value, $\mu^*$. The feasibility guarantee is weakened since Algorithm \ref{algorithm:greedy}
does not necessarily run to completion, but not by much if $\epsilon$ is sufficiently small.
In particular, we have the following alternative version of Theorem \ref{ratio}.

\paragraph{Theorem \ref{ratio} (Alternative)}
Suppose we have an instance of SCSC with $n\geq 1$.
Let $F$ be a function that is $\epsilon$-approximate to $f$.

Let $\mu^* > 0$ be given.
Suppose Algorithm \ref{algorithm:greedy} is run with input $F$, $c$, and $\tau$, but we exit the algorithm
at the first iteration $k$ such that $F_{\tau}(A_k)-F_{\tau}(A_{k-1}) \leq \mu^*$ and return $A=A_{k-1}$.
Then $f(A) \geq \tau - n((c_{max}/c_{min})\mu^*+2\epsilon)$ and if
$\epsilon < \mu^*/(4c_{max}\rho/c_{min}+2)$, then
\begin{align*}
  c(A) \leq& \frac{\rho}{1-\frac{4\epsilon c_{max} \rho}{c_{min} (\mu^*-2\epsilon)}}
             \left(\ln\left(\frac{\alpha}{\beta}\right) + 2\right)c(A^*)
\end{align*}
where $A^*$ is the optimal solution to the instance of SCSC.

\paragraph{Proof of Theorem \ref{ratio} (Alternative)}
Without loss of generality we re-define $f=\min\{f,\tau\}$ and $F=\min\{F,\tau\}$.
This way, $f=f_{\tau}$ and $F=F_{\tau}$.

First, we prove the feasibility. If Algorithm \ref{algorithm:greedy} runs to completion, then
the feasibility guarantee is clear since $f(A)\geq \tau - \epsilon$.
Suppose that Algorithm \ref{algorithm:greedy} did not run to completion, but instead returned
$A=A_{k-1}$ once $F(A_{k})-F(A_{k-1}) \leq \mu^*$. Let $x_k$ be the element that is selected in the
$k$th iteration.
Then for all $x$
\begin{align*}
  \Delta F(A, x) \leq \frac{c(x)}{c(x_k)}\Delta F(A, x_k) \leq \frac{c_{max}}{c_{min}}\mu^*.
\end{align*}
Therefore for all $x$ $\Delta f(A,x) \leq \Delta F(A, x) + 2\epsilon \leq (c_{max}/c_{min})\mu^* +2\epsilon$.
By the submodularity of $f$ we have that
\begin{align*}
  \tau - f(A) \leq f(S) - f(A) \leq \sum_{x}\Delta f(A,x) \leq n\left(\frac{c_{max}}{c_{min}}\mu^* +2\epsilon\right)
\end{align*}
and so the feasibility guarantee is proven.

The approximation ratio follows by using the same argument as in Theorem \ref{ratio} where
$\mu$ is replaced by $\mu^*-2\epsilon$ since
\begin{align*}
  \mu = \min\{f(A_i)-f(A_{i-1}): i\leq k\} \geq \min\{F(A_i)-F(A_{i-1}): i\leq k\}-2\epsilon > \mu^*-2\epsilon.
\end{align*}
\qed


%
%


\section{Proof of Theorem \ref{ratio}}
\label{appendix:theorem1}
The definitions and notation used in this section can be found in Section \ref{section:definitions} of the paper.

\paragraph{Theorem \ref{ratio}}
Suppose we have an instance of SCSC.
Let $F$ be a function that is $\epsilon$-approximate to $f$.

Suppose we run
Algorithm \ref{algorithm:greedy} with input $F$, $c$, and $\tau$.
Then $f(A) \geq \tau - \epsilon$.
And if $\mu > 4\epsilon c_{max}\rho/c_{min}$,
\begin{align*}
  c(A) \leq& \frac{\rho}{1-\frac{4\epsilon c_{max} \rho}{c_{min} \mu}}
             \left(\ln\left(\frac{\alpha}{\beta}\right) + 2\right)c(A^*)
\end{align*}
where $A^*$ is an optimal solution to the instance of SCSC.

\paragraph{Proof of Theorem \ref{ratio}}
\setcounter{equation}{0}
The feasibility guarantee is clear from the stopping condition on the greedy algorithm:
$f(A) \geq F(A) - \epsilon \geq \tau - \epsilon$.

We now prove the upper bound on $c(A)$ if $\mu > 4\epsilon c_{max}\rho/c_{min}$.
Without loss of generality we re-define $f=\min\{f,\tau\}$ and $F=\min\{F,\tau\}$.
This way, $f=f_{\tau}$ and $F=F_{\tau}$.
Notice that this
does not change that $F$ is an $\epsilon$-approximation of $f$ since the error is absolute.

Let $x_1,...,x_k$ be the elements of $A$ in the order that they were chosen
by the greedy algorithm. If $k=0$, then $c(A)=0$ and the approximation ratio is clear.
For the rest of the proof, we assume that $k\geq 1$.

We define a sequence of elements $\tilde{x}_1,...,\tilde{x}_k$ where
\begin{align*}
  \tilde{x}_i =
    \text{argmax}_{x \in S\setminus A_{i-1}}\frac{\Delta f(A_{i-1}, x)}{c(x)}.
\end{align*}
$\tilde{x}_i$ has the most cost-effective marginal gain of being added to $A_{i-1}$
according to $f$, while $x_i$ has the most cost-effective marginal gain of being added to
$A_{i-1}$ according to $F$.
Note that the same element can appear multiple times in the sequence
$\tilde{x}_1,...,\tilde{x}_k$. In addition, we have a lower bound on $\Delta f(A_{i-1}, \tilde{x}_i)$:
\begin{align}
  \Delta f(A_{i-1}, \tilde{x}_i)
  \geq \frac{c(\tilde{x}_i)}{c(x_i)}\Delta f(A_{i-1}, x_i)
  \geq \frac{c_{min}}{c_{max}}\mu \label{eqn2}.
\end{align}

Our argument to bound $c(A)$ will follow
the following three steps:
\textbf{(a)} We bound $c(A)$ in terms of the costs of the elements
$\tilde{x}_1,...,\tilde{x}_k$.
\textbf{(b)} We charge the elements
of $A^*$ with the costs of the elements $\tilde{x}_1,...,\tilde{x}_k$,
and bound $c(A)$ in terms of the total charge on all elements in $A^*$.
\textbf{(c)} We bound the total charge on the elements of $A^*$ in terms of $c(A^*)$.

\paragraph{(a)} First, we bound $c(A)$ in terms of the costs of the elements
$\tilde{x}_1,...,\tilde{x}_k$.
At iteration $i$ of Algorithm \ref{algorithm:greedy}, the most cost-effective element
to add to the set $A_{i-1}$ according to $F$ is $x_i$.
Using the fact that $F$ is $\epsilon$-approximate to $f$, we can bound
how much more cost-effective $\tilde{x}_i$ is compared to $x_i$ according to $f$
as follows:
\begin{align*}
  \frac{\Delta f(A_{i-1},x_i) + 2\epsilon}{c(x_i)} \geq
  \frac{\Delta F(A_{i-1},x_i)}{c(x_i)} \geq
  \frac{\Delta F(A_{i-1},\tilde{x}_i)}{c(\tilde{x}_i)} \geq
  \frac{\Delta f(A_{i-1},\tilde{x}_i) - 2\epsilon}{c(\tilde{x}_i)}
\end{align*}
which implies that
\begin{align}
  \frac{\Delta f(A_{i-1}, x_i)}{c(x_i)} +
     \frac{2\epsilon(c(x_i) + c(\tilde{x}_i))}{c(x_i)c(\tilde{x}_i)}
     \geq \frac{\Delta f(A_{i-1}, \tilde{x}_i)}{c(\tilde{x}_i)}. \label{eqn3}
\end{align}
$\Delta f(A_{i-1}, x_i) \geq \mu > 0$ by assumption, and
$\Delta f(A_{i-1}, \tilde{x}_i) \geq (c_{min}/c_{max})\mu > 0$ by
Equation (\ref{eqn2}). Therefore we can re-arrange Equation (\ref{eqn3}) to be
\begin{align}
  \frac{c(\tilde{x}_i)}{\Delta f(A_{i-1}, \tilde{x}_i)}
  + \alpha_i
  \geq \frac{c(x_i)}{\Delta f(A_{i-1}, x_i)} \text{, where }
  \alpha_i = \frac{2\epsilon(c(x_i)+c(\tilde{x}_i))}{\Delta f(A_{i-1}, \tilde{x}_i)\Delta f(A_{i-1}, x_i)}. \label{eqn4}
\end{align}

Inequality (\ref{eqn4}) and the submodularity of $c$ imply that
\begin{align}
 c(A) &\leq \sum_{i=1}^k c(x_i)
      = \sum_{i=1}^k \Delta f(A_{i-1},x_i)
         \frac{c(x_i)}{\Delta f(A_{i-1},x_i)}\nonumber \\
      &\leq \sum_{i=1}^k \Delta f(A_{i-1}, {x_i})
          \left(\frac{c(\tilde{x}_i)}{\Delta f(A_{i-1}, \tilde{x}_i)}
          + \alpha_i\right) \nonumber \\
      &= \sum_{i=1}^k \Delta f(A_{i-1},x_i)\frac{c(\tilde{x}_i)}{\Delta f(A_{i-1}, \tilde{x}_i)}
            + \sum_{i=1}^k \frac{2\epsilon(c(\tilde{x}_i) + c(x_i))}
              {\Delta f(A_{i-1},\tilde{x}_i)}. \label{eqn5}
\end{align}

We now bound the second term on the right side of Equation (\ref{eqn5}) by
\begin{align}
  \sum_{i=1}^k \frac{2\epsilon(c(\tilde{x}_i) + c(x_i))}
    {\Delta f(A_{i-1},\tilde{x}_i)}
  &= \sum_{i=1}^k \frac{2\epsilon c(\tilde{x}_i)}
    {\Delta f(A_{i-1},\tilde{x}_i)}
    + \sum_{i=1}^k \frac{2\epsilon c(x_i)}{\Delta f(A_{i-1},\tilde{x}_i)}\nonumber \\
  &\leq \sum_{i=1}^k \frac{2\epsilon c(x_i)}{\Delta f(A_{i-1},x_i)}
      + \sum_{i=1}^k \frac{2\epsilon c(x_i)}{\Delta f(A_{i-1},\tilde{x}_i)} \nonumber \\
  &\leq \frac{4\epsilon c_{max}}{c_{min}\mu}\sum_{i=1}^k c(x_i)\nonumber \\
  &\leq \frac{4\epsilon c_{max}\rho}{c_{min}\mu}c(A).
  \label{eqn6}
\end{align}
The second to last inequality in Equation (\ref{eqn6}) follows from the fact that
$\Delta f(A_{i-1},x_i) \geq \mu \geq (c_{min}/c_{max})\mu$, and that by Equation (\ref{eqn2})
$\Delta f(A_{i-1},\tilde{x}_i) \geq (c_{min}/c_{max})\mu$.
The last inequality in Equation (\ref{eqn6}) uses the definition of the curvature $\rho$ of $c$.

Combining Equations (\ref{eqn5}) and (\ref{eqn6}) gives us the following bound on $c(A)$
in terms of the costs of the elements $\tilde{x}_1,...,\tilde{x}_k$:
\begin{align}
  \left( 1 - \frac{4\epsilon c_{max}\rho}{c_{min}\mu} \right)c(A)
  \leq \sum_{i=1}^k \frac{\Delta f(A_{i-1},x_i)}{\Delta f(A_{i-1},\tilde{x}_i)} c(\tilde{x}_i). \label{eqn7}
\end{align}

\paragraph{(b)} Next, we charge the elements
of $A^*$ with the costs of the elements $\tilde{x}_1,...,\tilde{x}_k$,
and bound $c(A)$ in terms of the total charge on all elements in $A^*$.
By this we mean that we give each $y\in A^*$ a portion of the total cost of
the elements $\tilde{x}_1,...,\tilde{x}_k$. In particular,
we give each $y\in A^*$ a charge of $w(y)$, defined by
\begin{align*}
  w(y) = \sum_{i=1}^k(\pi_i(y) - \pi_{i+1}(y))\omega_i
  \text{, where }
  \omega_i = \frac{c(\tilde{x}_i)}{\Delta f(A_{i-1}, \tilde{x}_i)},
  \text{ and }
  \pi_i(y) = \begin{cases}
        \Delta f(A_{i - 1}, y) & i \in \{1,...,k\} \\
        \Delta f(A, y) & i = k+1
     \end{cases}.
\end{align*}
Recall that $\Delta f(A_{i-1}, \tilde{x}_i) > 0$ for all $i$ by Equation (\ref{eqn2}),
and so we can define $\omega_i$ as above. \citeauthor{wan2010} charged the elements of $A^*$ with the
cost of elements $x_1,...,x_k$ analogously to the above. We charge with the cost of elements
$\tilde{x}_1,...,\tilde{x}_k$ because they exhibit diminishing cost-effectiveness, i.e.
$\omega_i-\omega_{i-1}\geq 0$ for all $i\in\{1,...,k\}$, which is needed to proceed with the argument.
Because we choose $x_1,...,x_k$ with
$F$, which is not monotone submodular, $x_1,...,x_k$ do not exhibit diminishing cost-effectiveness
even if we replace $f$ with $F$ in the definition of $w(y)$ above. We now follow an argument analogous to
\citeauthor{wan2010} but with the elements $\tilde{x}_1,...,\tilde{x}_k$ in order to prove Equation (\ref{eqn10}).

Consider any $y\in A^*$. We can re-write $w(y)$ as
\begin{align*}
  w(y) & = \sum_{i=1}^{k}(\pi_i(y) - \pi_{i+1}(y))\omega_i\\
        &= \sum_{i=1}^{k}\pi_i(y)\omega_i - \sum_{i=1}^{k}\pi_{i+1}(y)\omega_i\\
        &= \pi_1(y)\omega_1 + \sum_{i=2}^{k}\pi_i(y)\omega_i
         - \sum_{i=2}^{k}\pi_i(y)\omega_{i-1} - \pi_{k+1}(y)\omega_{k}\\
        &= \pi_1(y)\omega_1 - \Delta f(A,y)\omega_k
        + \sum_{i=2}^{k}(\omega_i - \omega_{i-1})\pi_i(y).
\end{align*}
Summing over $y\in A^*$, we have
\begin{align}
  \sum_{y\in A^*}w(y)
  &= \omega_1\sum_{y\in A^*}\pi_1(y)
    + \sum_{i=2}^{k}(\omega_i - \omega_{i-1})\sum_{y\in A^*}\pi_i(y) - \sum_{y\in A^*}\Delta f(A,y)\omega_k. \label{eqn6.5}
\end{align}

On the other hand, starting with Equation (\ref{eqn7}), we see that
\begin{align}
  \left(1 - \frac{4\epsilon c_{max}\rho}{\mu c_{min}}\right)c(A)
  & \leq \sum_{i=1}^k \frac{\Delta f(A_{i-1},x_i)}{\Delta f(A_{i-1},\tilde{x}_i)} c(\tilde{x}_i)\nonumber \\
  &= \Delta f(A_{k-1},x_k)\omega_k
     + \sum_{i=1}^{k-1} \Delta f(A_{i-1},x_i)\omega_i \nonumber \\
  &= \Delta f(A_{k-1},x_k)\omega_k
    + \sum_{i=1}^{k-1} (\sum_{j=i}^k \Delta f(A_{j-1},x_j)
    - \sum_{j=i+1}^k \Delta f(A_{j-1},x_j))\omega_i \nonumber \\
  &= \sum_{i=1}^k\sum_{j=i}^k \Delta f(A_{j-1},x_j)\omega_i
    - \sum_{i=1}^{k-1}\sum_{j=i+1}^k \Delta f(A_{j-1},x_j)\omega_i \nonumber \\
  &= \omega_1\sum_{j=1}^k \Delta f(A_{j-1},x_j)
    + \sum_{i=2}^k\sum_{j=i}^k \Delta f(A_{j-1},x_j)\omega_i
    - \sum_{i=2}^{k}\sum_{j=i}^k \Delta f(A_{j-1},x_j)\omega_{i-1} \nonumber \\
  &= \omega_1\sum_{j=1}^k \Delta f(A_{j-1},x_j)
    + \sum_{i=2}^k(\omega_i - \omega_{i-1})\sum_{j=i}^k \Delta f(A_{j-1},x_j). \label{eqn8.5}
\end{align}

In order to find a link between Equations (\ref{eqn6.5}) and (\ref{eqn8.5}), we first notice that
for any $i\in\{1,...,k\}$
\begin{align*}
  \sum_{j=i}^k \Delta f(A_{j-1},x_j)
  = f(A_k) - f(A_{i-1})
  \leq \sum_{y\in A^*} \Delta f(A_{i-1},y)
  = \sum_{y\in A^*} \pi_i(y).
\end{align*}
In addition, for any $i\in\{1,...,k\}$
\begin{align*}
  \omega_i - \omega_{i-1}
  &= \frac{c(\tilde{x}_i)}{\Delta f(A_{i-1},\tilde{x}_i)} - \frac{c(\tilde{x}_{i-1})}{\Delta f(A_{i-2},\tilde{x}_{i-1})} \\
  &\geq \frac{c(\tilde{x}_i)}{\Delta f(A_{i-2},\tilde{x}_i)}
    - \frac{c(\tilde{x}_{i-1})}{\Delta f(A_{i-2},\tilde{x}_{i-1})}\\
  &\geq \frac{c(\tilde{x}_{i-1})}{\Delta f(A_{i-2},\tilde{x}_{i-1})}
    - \frac{c(\tilde{x}_{i-1})}{\Delta f(A_{i-2},\tilde{x}_{i-1})}
  = 0.
\end{align*}

We may therefore link Equations (\ref{eqn6.5}) and (\ref{eqn8.5}) to see that
\begin{align}
  \left(1 - \frac{4\epsilon c_{max}\rho}{\mu c_{min}}\right)c(A)
  &\leq \omega_1\sum_{j=1}^k \Delta f(A_{j-1},x_j)
    + \sum_{i=2}^k(\omega_i - \omega_{i-1})\sum_{j=i}^k \Delta f(A_{j-1},x_j) \nonumber\\
  &\leq \sum_{y\in A^*}w(y) + \sum_{y\in A^*}\Delta f(A,y)\omega_k. \label{eqn10}
\end{align}

Consider $y\in A^*$.
$f(A)$ is not necessarily $\tau$ since the stopping condition for Algorithm \ref{algorithm:greedy}
is only that $F(A)\geq\tau$. In this case, if $\Delta f(A,y)\neq 0$ for $y\in A^*$ (which implies that
$y\notin A_{k-1}$)
then by the submodularity of $f$
\begin{align*}
  \Delta f(A,y)\omega_k
  = \Delta f(A,y)\frac{c(\tilde{x}_k)}{\Delta f(A_{k-1},\tilde{x}_k)}
  \leq \Delta f(A,y)\frac{c(y)}{\Delta f(A_{k-1},y)}
  \leq \Delta f(A,y)\frac{c(y)}{\Delta f(A,y)}
  = c(y).
\end{align*}
Therefore we can bound $c(A)$ in terms of the total charge on all
elements in $A^*$:
\begin{align}
  \left(1 - \frac{4\epsilon c_{max}\rho}{\mu c_{min}}\right)c(A)
  \leq \sum_{y\in A^*}w(y) + \sum_{y\in A^*}c(y)
  \leq \sum_{y\in A^*}w(y) + \rho c(A^*). \label{eqn11}
\end{align}

\paragraph{(c)} Finally, we bound the total charge on the elements of $A^*$ in terms of $c(A^*)$.

We first define a value $\ell_y$ for every $y\in A^*$. For each $y\in A^*$,
if $\pi_1(y) = 0$ we set $\ell_y=0$,
otherwise $\ell_y$ is the value in $\{1,...,k\}$ such that if $i\in \{1,...,\ell_y\}$
then $\pi_i(y) > 0$, and if $i\in \{\ell_y+1,...,k\}$ then $\pi_i(y) = 0$. Such an $\ell_y$
can be set since $f$ is submodular and monotonic.
Then
\begin{align}
  w(y) &= \sum_{i=1}^{\ell_y}(\pi_i(y) - \pi_{i+1}(y))\omega_i\nonumber\\
       &\leq \sum_{i=1}^{\ell_y}(\pi_i(y) - \pi_{i+1}(y))\frac{c(y)}{\pi_i(y)}\nonumber\\
       &\leq c(y)\left(\left(\sum_{i=1}^{\ell_y-1}\frac{\pi_i(y) - \pi_{i+1}(y)}{\pi_i(y)}\right)+ 1\right) \nonumber \\
       &\leq c(y)\left( \sum_{i=1}^{\ell_y - 1}(\ln(\pi_i(y))-\ln(\pi_{i+1}(y))) + 1\right)\nonumber\\
       &= c(y)\left( \ln(\pi_1(y))-\ln(\pi_{\ell_y}(y)) + 1\right) \nonumber \\
       &\leq c(y)\left(\ln\left(\frac{\alpha}{\beta}\right) + 1\right). \label{eqn12}
\end{align}
The third to last inequality follows since
\begin{align*}
  \frac{\pi_i(y) - \pi_{i+1}(y)}{\pi_i(y)}
  = \int_{\pi_{i+1}(y)}^{\pi_i(y)} \frac{1}{\pi_i(y)}dx \leq \int_{\pi_{i+1}(y)}^{\pi_i(y)}\frac{1}{x}dx
  = \ln(\pi_i(y))-\ln(\pi_{i+1}(y)).
\end{align*}

We sum inequality (\ref{eqn12}) over all $y \in A^*$ to get that
\begin{align}
  \sum_{y\in A^*}w(y) \leq \left(\ln\left(\frac{\alpha}{\beta}\right) + 1\right)\sum_{y\in A^*}c(y)
  \leq \rho\left(\ln\left(\frac{\alpha}{\beta}\right) + 1\right)c(A^*). \label{eqn13}
\end{align}
Finally, we combine inequality (\ref{eqn11}) and inequality (\ref{eqn13}) to see that
\begin{align*}
  \left(1 - \frac{4\epsilon c_{max}\rho}{c_{min}\mu}\right)c(A) &\leq \sum_{y\in A^*}w(y) + \rho c(A^*)
       \leq \rho\left(\ln\left(\frac{\alpha}{\beta}\right)+1\right)c(A^*) + \rho c(A^*).
\end{align*}
If $\mu > (4\epsilon c_{max} \rho)/(c_{min})$, this completes the proof of the approximation
guarantee in the theorem statement. \qed

\section{Proof of Theorem \ref{ratio2}}
\label{appendix:theorem2}
The definitions and notation used in this section can be found in Section \ref{section:ratios} of the paper.

\paragraph{Theorem \ref{ratio2}}
Suppose we have an instance of SCSC.
Let $F$ be a function that is $\epsilon$-approximate to $f$.

Suppose we run
Algorithm \ref{algorithm:greedy} with input $F$, $c$, and $\tau$.
Then $f(A) \geq \tau - \epsilon$.
And if $\mu > 4\epsilon c_{max}\rho/c_{min}$, then for any
$\gamma \in \left(0,1-4\epsilon c_{max}\rho/c_{min}\mu\right)$,
\begin{align*}
  c(A) \leq& \frac{\rho}{1-\frac{4\epsilon c_{max} \rho }{c_{min} \mu} - \gamma}
             \left(\ln\left(\frac{n\alpha\rho}{\gamma\mu}\right) + 2\right)c(A^*)
\end{align*}
where $A^*$ is an optimal solution to the instance of SCSC.

\paragraph{Proof of Theorem \ref{ratio2}}
\label{appendix:theorem2}

\setcounter{equation}{0}
The argument for the proof of Theorem \ref{ratio2} is the same as Theorem \ref{ratio},
except for part \textbf{(c)} which is what we present here.
In particular, we have gotten to the point of the proof of Theorem \ref{ratio} where we have proven that
\begin{align}
  \left(1-\frac{4\epsilon c_{max}\rho}{c_{min}\mu}\right)c(A) \leq
    \sum_{y\in A^*}w(y) + \rho c(A^*). \label{eqn13}
\end{align}

Let $\lambda > 0$. We first define a value $m_y$ for every $y\in A^*$. For each $y\in A^*$,
if $\pi_1(y) \leq \lambda$ we set $m_y=0$,
otherwise $m_y$ is the value in $\{1,...,k\}$ such that if $i\in\{1,...,m_y\}$ then
$\pi_i(y) > \lambda$, and if $i\in\{m_y+1,...,k\}$ then $\pi_i(y)\leq\lambda$.
Such an $m_y$ can be set since $f$ is submodular and monotonic. Then
\begin{align}
  w(y)= \sum_{i=1}^{m_y}(\pi_i(y)-\pi_{i+1}(y))\omega_i
   + \sum_{i=m_y+1}^{k}(\pi_i(y)-\pi_{i+1}(y))\omega_i. \label{eqn14}
\end{align}

A similar analysis as in the proof of Theorem \ref{ratio} can be used to show that
\begin{align}
  \sum_{i=1}^{m_y}(\pi_i(y)-\pi_{i+1}(y))\omega_i
  \leq c(y)\left(\ln\left(\frac{\alpha}{\lambda}\right) + 1\right).\label{eqn15}
\end{align}
The remaining part of inequality (\ref{eqn14}) can be bounded by
\begin{align}
  \sum_{i=m_y+1}^{k}(\pi_i(y)-\pi_{i+1}(y))\omega_i
  &\leq \sum_{i=m_y+1}^{k}(\pi_i(y)-\pi_{i+1}(y))\frac{c(x_i)}{\Delta f(A_{i-1},x_i)}\nonumber \\
  &\leq \sum_{i=m_y+1}^{k}\pi_i(y)\frac{c(x_i)}{\Delta f(A_{i-1},x_i)} \nonumber \\
  &< \sum_{i=m_y+1}^{k}\lambda\frac{c(x_i)}{\mu}\nonumber \\
  &\leq \frac{\lambda\rho}{\mu}c(A).\label{eqn16}
\end{align}

Summing Equation (\ref{eqn14}) over all $y\in A^*$ and applying the upper bounds in Equations
(\ref{eqn15}) and (\ref{eqn16}) gives us that
\begin{align}
  \sum_{y\in A^*}w(y) &\leq \rho\left(\ln\left(\frac{\alpha}{\lambda}\right)+1\right)c(A^*)
      + \frac{\lambda\rho n}{\mu}c(A) \label{eqn17}
\end{align}
By combining Equations (\ref{eqn17}) and (\ref{eqn13}), we have that
\begin{align*}
  \left(1 - \frac{4\epsilon c_{max}\rho}{c_{min}\mu} - \frac{\lambda\rho n}{\mu}\right)c(A)
      &\leq \rho\left(\ln\left(\frac{\alpha}{\lambda}\right)+2\right)c(A^*).
\end{align*}

If we set $\lambda = (\gamma\mu)/(n\rho)$ we have the approximation ratio in
the theorem statement.
$\qed$

\section{Application and Experiments}
\label{appendix:influence}

\paragraph{Influence Threshold (IT) (Non-Simulation)}
In contrast to the version of IT in Section \ref{section:experiments} of the paper, we may define
IT to directly use the influence model as follows.

Let $G=(V,E)$ be a social network where nodes $V$ represents users,
edges $E$ represent social connections, and $\mathcal{D}$ is a probability distribution
over subsets of $E$ that represents the probability of a set of edges being ``alive''.
Activation of users in the social network starts from an
initial seed set and then propagates across alive edges. For $X\subseteq S$,
$f(X)$ is the expected number of reachable nodes in $V$ from $X$ when a set of alive edges is
sampled from $\mathcal{D}$.
$c:2^V\to\mathbb{R}_{\geq 0}$ is a monotone submodular function that gives the cost of
seeding a set of users. The Influence Threshold problem (IT) is to find a seed set $A$ such that $c(A)$ is minimized and
$f(A)\geq\tau$.

\paragraph{$f$ as Average Reachability}
One popular choice for the distribution $\mathcal{D}$ is the Independent Cascade (IC) model \cite{kempe2003}.
In the IC model, every edge has a probability assigned to it $w_E:E\to [0,1]$. Each edge $e$ is independently
alive with probability $w_E(e)$. However, computing the expected influence $f$ under the IC model is $\#P$-hard and
therefore impractical to compute \cite{chen2010}.

As an alternative to working directly with the influence model
\citeauthor{kempe2003} proposed a simulation-based approach to approximating it:
Random samples from $\mathcal{D}$ are drawn,
and for every $X\subseteq S$ $f(X)$ is approximated by the average number of reachable nodes from $X$
over the samples.
The resulting approximation of $f$ is unbiased, converges to the expected value, and is monotone submodular.
This simulation-based approach is also advantageous since it can be used for more complex models than IC or
for instances generated from traces \cite{cohen2014}.

Because the average reachability is monotone submodular,
it is easy to translate approximation results for IT where $f$ is replaced
by the average reachability into approximation results for IT (see Proposition \ref{translate} of Appendix
\ref{section:submodularF}).

\paragraph{Additional Experimental Setup Details}
We use two real social networks: the Facebook ego network \cite{leskovec2012} with $n=4039$,
and the ArXiV General Relativity collaboration
network with $n=5242$ \cite{leskovec2007}, which we refer to as GrQc.

Influence propagation follows the Independent Cascade (IC) model \cite{kempe2003}.
The probabilities assigned to the edges $w_E:E\to[0,1]$ follow the
weighted cascade model \cite{kempe2003}. In the weighted cascade model,
an edge that goes from $u\in V$ to $v\in V$
is assigned probability $\frac{q}{d_v}$ where $d_v$
is the number of incoming edges to node $v$ and $q\in (0,1]$. For facebook $q=0.5$, and
for GrQc $q=0.8$.

The average reachability oracle, $f$, is over $N=25000$ random realizations of the influence graph.
The approximate average reachability oracle of \citeauthor{cohen2014}, $F$, is computed over these
realizations with various oracle errors $\epsilon$ and the greedy algorithm is run using these oracles.
For comparison, we also run the greedy algorithm with $f$.

For the cost function $c:2^V\to\mathbb{R}_{\geq 0}$, we
choose a cost $c_v$ for each $v\in V$ by sampling from a normal distribution with mean 1 and
standard deviation 0.1, and then define $c(X)=\sum_{x\in X}c_x$.

To select the parameter $\gamma$ when computing the ratio of Theorem \ref{ratio2},
we discretize the domain of $\gamma$, compute
the ratio on each of the points, and select the $\gamma$ that gives the smallest approximation ratio.

\paragraph{Additional Experimental Results}
\begin{figure*}
  \centering
  \hspace{-10pt}
  \subfigure[][facebook, r1, $\rho=1.0$] {
    \label{fig5}
    \includegraphics[width=0.24\textwidth]{Images/newplots/facebook_ratios1}
  }
  \hspace{-10pt}
  \subfigure[][facebook, r1, $\rho=1.2$] {
    \label{fig6}
    \includegraphics[width=0.24\textwidth]{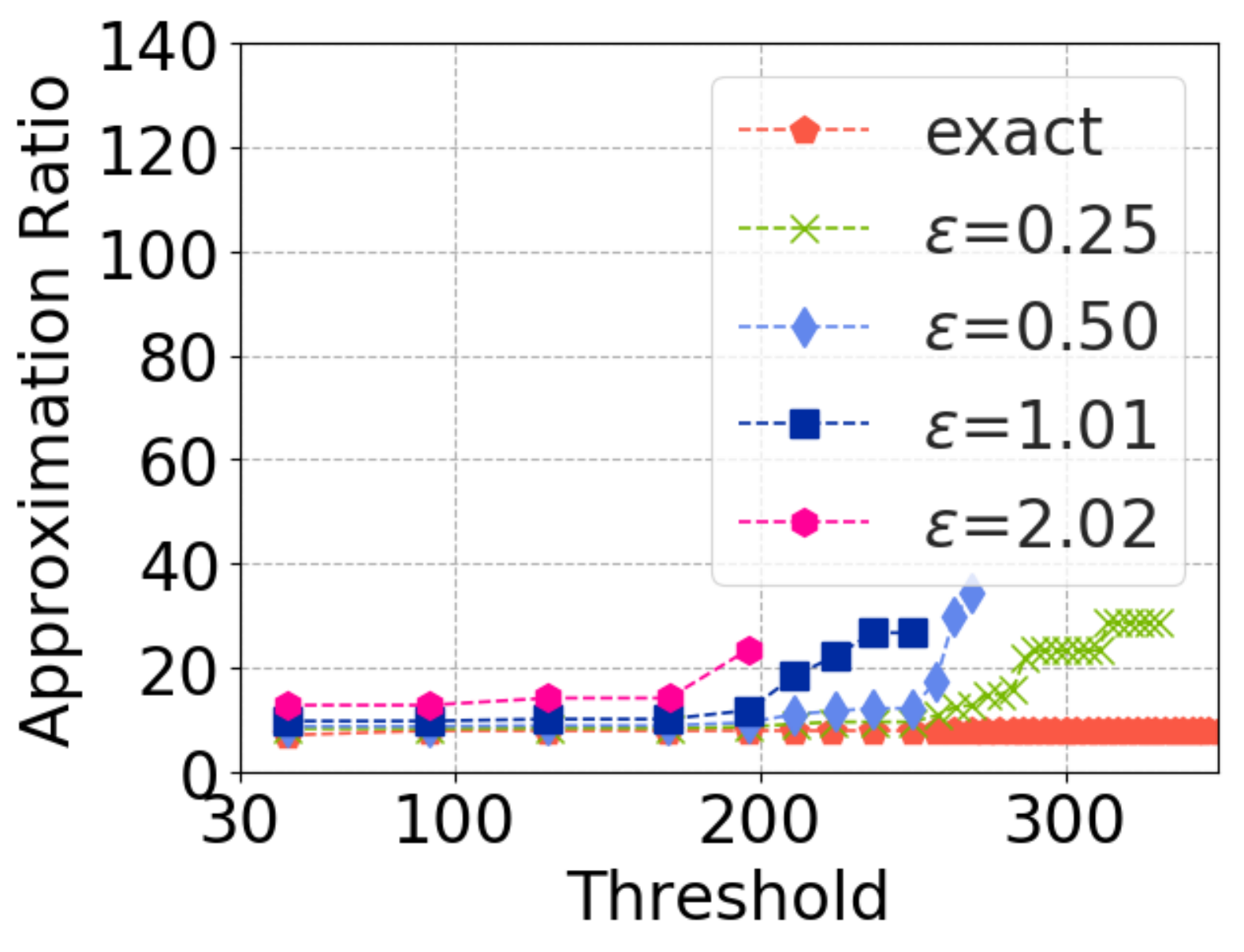}
  }
  \hspace{-10pt}
  \subfigure[][facebook, r1, $\rho=1.4$] {
    \label{fig7}
    \includegraphics[width=0.24\textwidth]{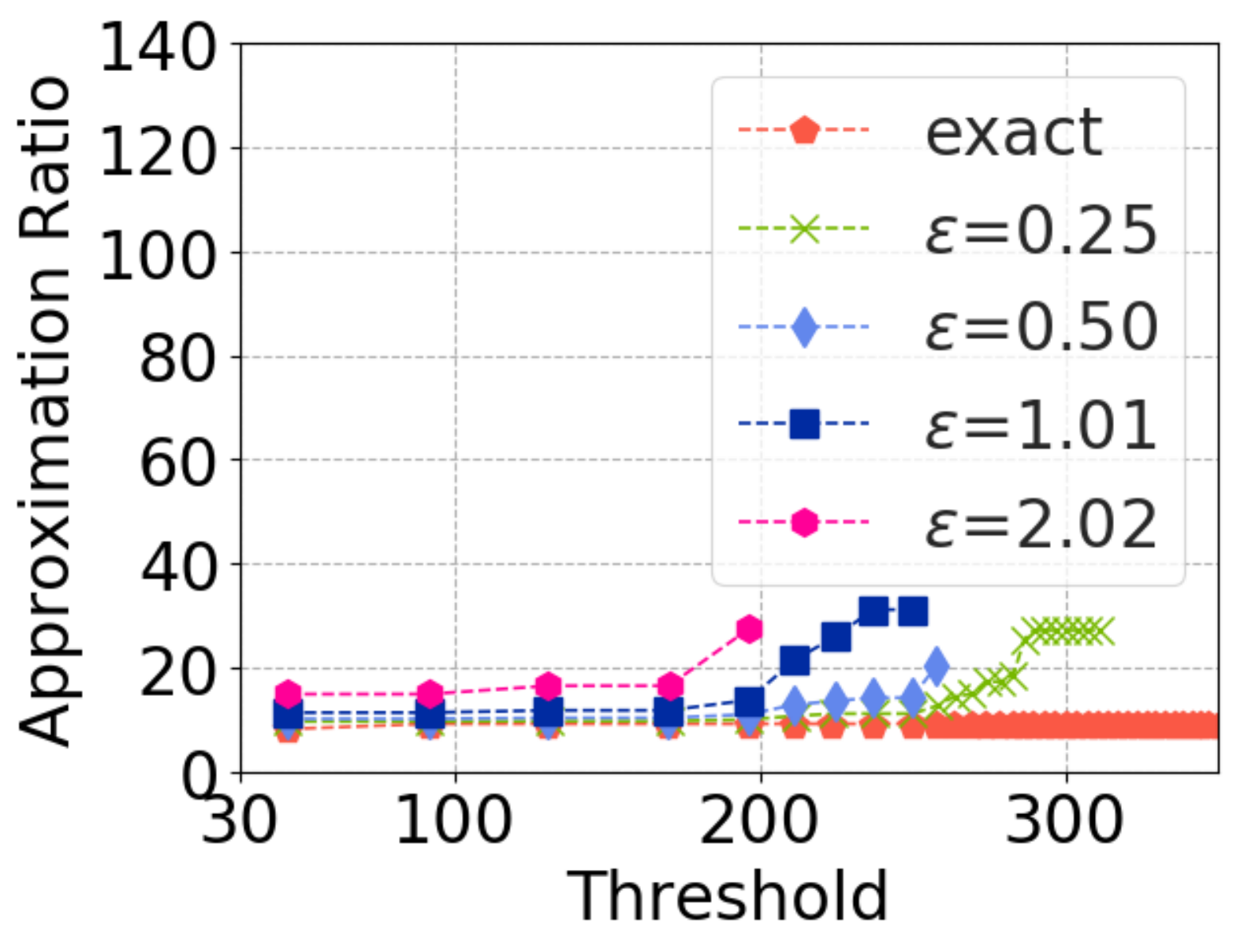}
  }
  \hspace{-10pt}
  \subfigure[][facebook, r1, $\rho=1.6$] {
    \label{fig8}
    \includegraphics[width=0.24\textwidth]{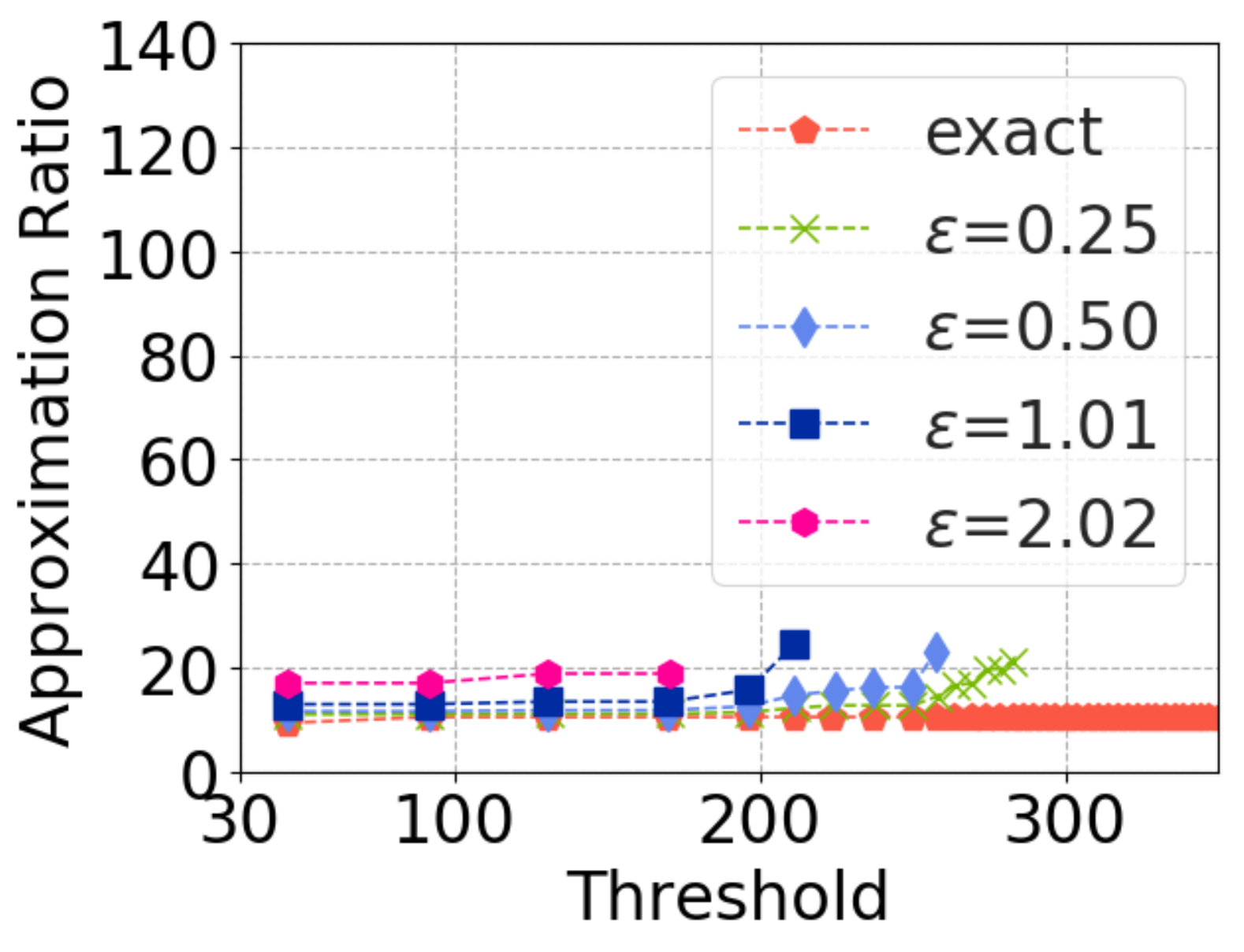}
  }
  \hspace{-10pt}
  \subfigure[][facebook, r2, $\rho=1.0$] {
    \label{fig9}
    \includegraphics[width=0.24\textwidth]{Images/newplots/facebook_ratios2}
  }
  \hspace{-10pt}
  \subfigure[][facebook, r2, $\rho=1.2$] {
    \label{fig10}
    \includegraphics[width=0.24\textwidth]{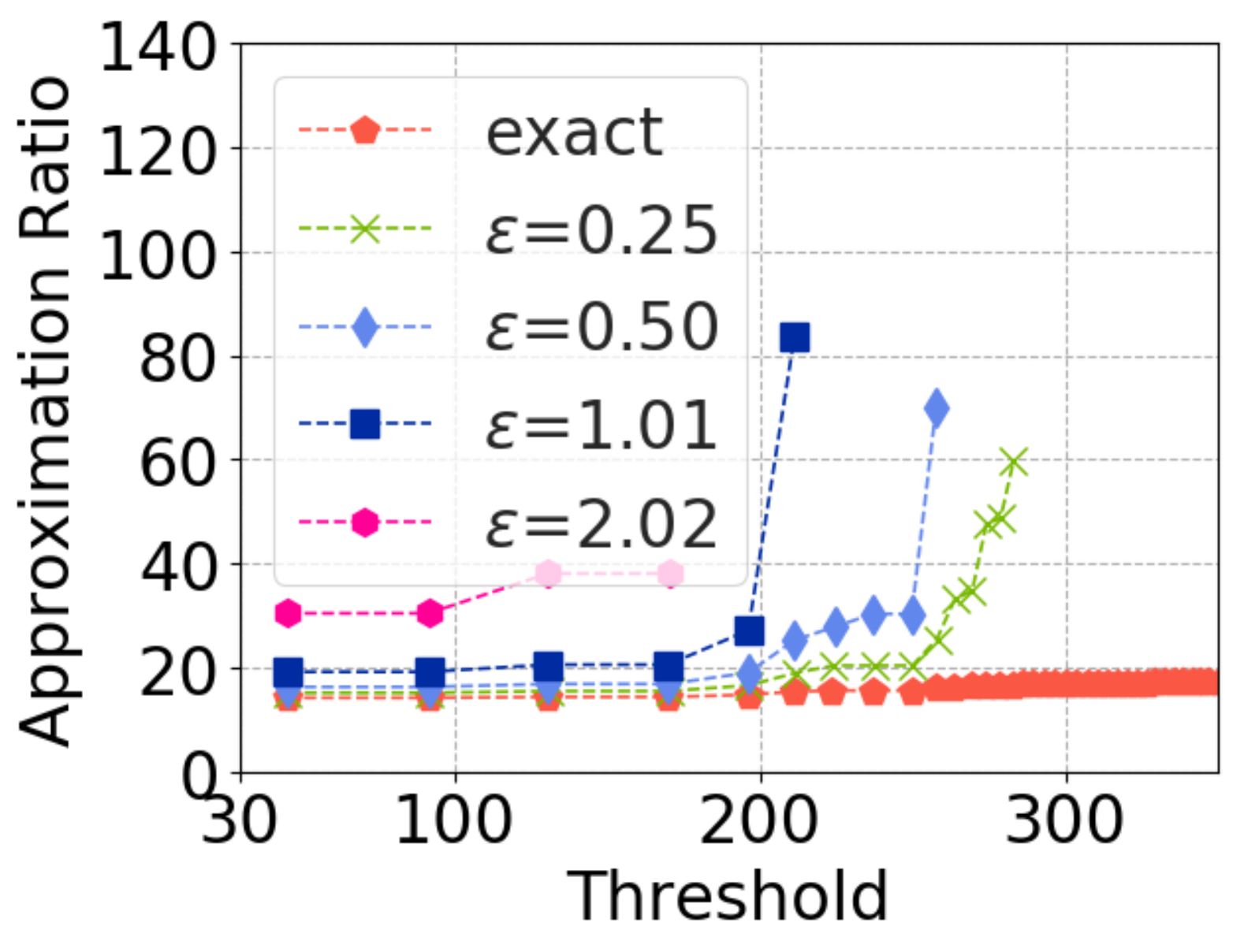}
  }
  \hspace{-10pt}
  \subfigure[][facebook, r2, $\rho=1.4$] {
    \label{fig11}
    \includegraphics[width=0.24\textwidth]{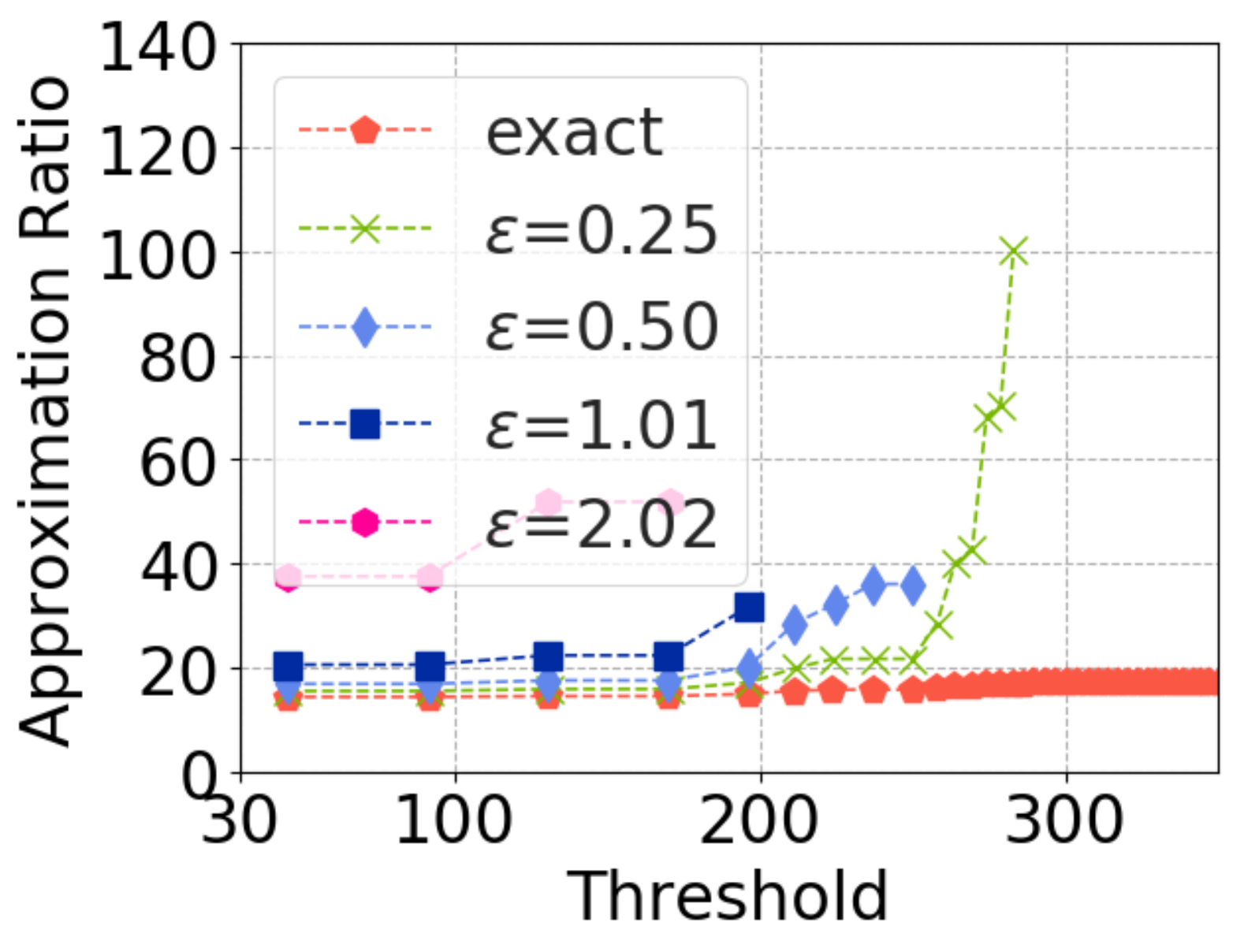}
  }
  \hspace{-10pt}
  \subfigure[][facebook, r2, $\rho=1.6$] {
    \label{fig12}
    \includegraphics[width=0.24\textwidth]{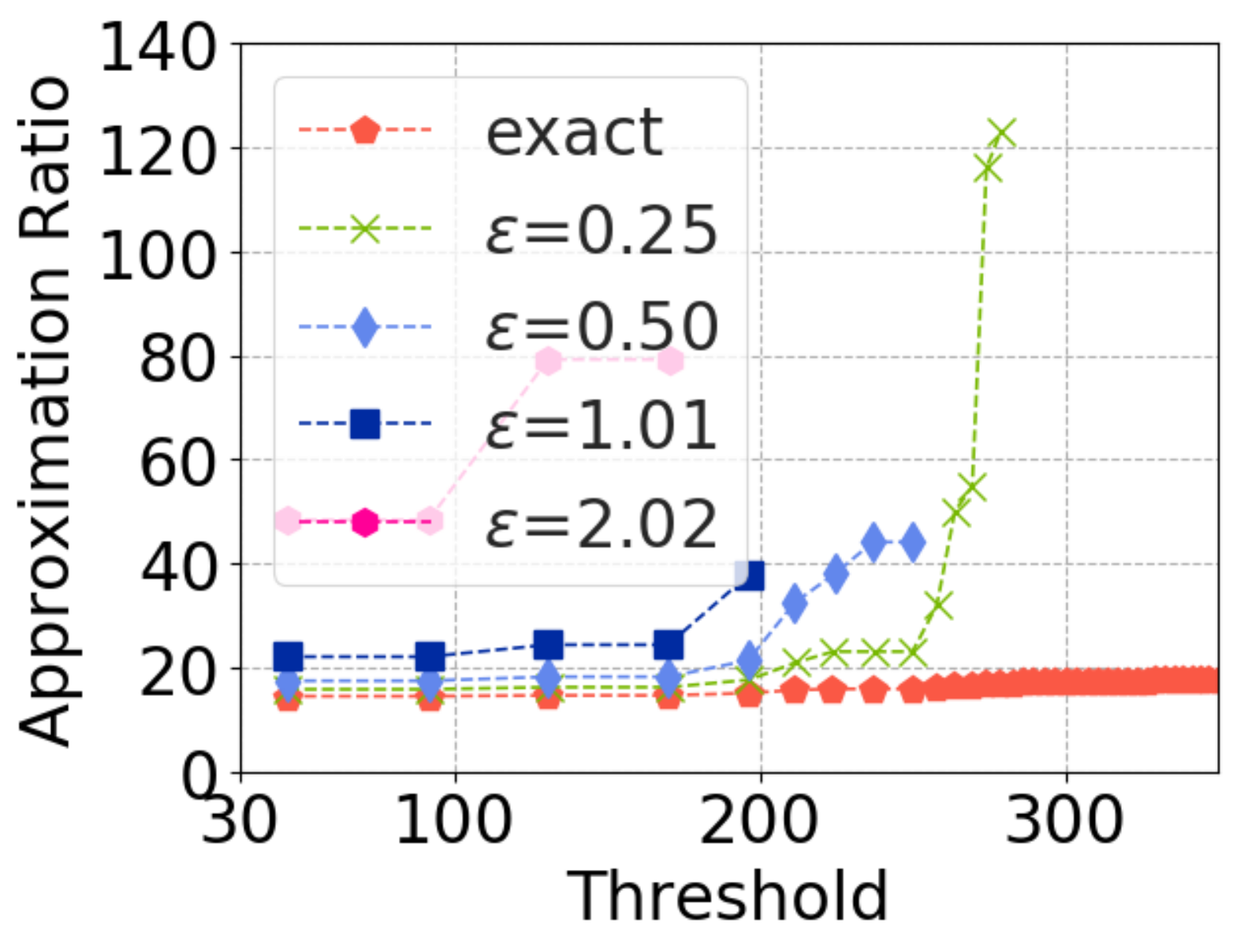}
  }
  \caption{The approximation ratios of Theorem \ref{ratio} (r1) and an upper bound on that of Theorem \ref{ratio2} (r2)
  at thresholds indicated by the markers.}
\end{figure*}


The experimental results presented in Section \ref{section:experiments} of the paper consider a
modular cost function. However, the approximation results presented in Theorems \ref{ratio} and
\ref{ratio2} are for general submodular cost functions. In this section, we present additional experimental results
analogous to those in Section \ref{section:experiments} of the paper but with non-modular cost function.

Recall from Section \ref{section:definitions} that a measure of how far a function is from being is
modular is measured by its curvature, $\rho \in [1,\infty)$. Our approximation ratios in Theorems
\ref{ratio} and \ref{ratio2} depend on $\rho$: the smaller $\rho$, and hence the closer to being modular
the function is, the better the approximation ratio.

Notice that the greedy algorithm, presented in Section \ref{section:definitions}, chooses elements only according
to singleton costs. Therefore, the experimental results in Section \ref{section:experiments} of the paper
can easily be extended to cost functions that
are not modular by simply computing the ratio for different values of cost curvature, $\rho$; higher values
of $\rho$ would require smaller epsilon to get the same ratio.

The approximation ratio of Theorem \ref{ratio} on the Facebook dataset is plotted in
Figures \ref{fig5} to \ref{fig8} for varying curvature $\rho$, and that of Theorem \ref{ratio2}
is plotted in Figures \ref{fig9} to \ref{fig12}. As $\rho$ increases, the approximation ratios are
subtly greater. The greater that $\rho$ is, the greater $\mu$ must be relative to $\epsilon$ in order
to have the approximation ratios of Theorems \ref{ratio} and \ref{ratio2}. This results in the ratios
of Theorems \ref{ratio} and \ref{ratio2} not being guaranteed at lower thresholds for larger $\rho$.

\end{document}